\documentclass[useAMS,usenatbib]{mn2e}
\usepackage{graphicx}
\usepackage{subfigure}
\usepackage{amsmath}
\usepackage{cancel}
\usepackage{color}

\title[Signatures of cloud--cloud collisions]{Isolating signatures of major cloud--cloud collisions using position--velocity diagrams}

\author[T. J. Haworth et al.]
{\parbox{\textwidth}{T. J. Haworth$^{1}$\thanks{E-mail: \texttt{thaworth@ast.cam.ac.uk}},
E. J. Tasker$^2$,
Y. Fukui$^3$,
K. Torii$^3$,
J. E. Dale$^{4}$,
K. Shima$^{2}$,
\\K. Takahira$^2$,
A. Habe$^2$
and K. Hasegawa$^3$
}\vspace{0.4cm}\\
\parbox{\textwidth}{$^{1}$Institute of Astronomy, Madingley Rd, Cambridge, CB3 0HA, UK\\
$^{2}$ Department of Physics, Faculty of Science, Hokkaido University, Kita-ku, Sapporo 060-0810, Japan\\
$^{3}$ Department of Physics and Astrophysics, Nagoya University, Chikusa-ku, Nagoya, Aichi 464-8601, Japan\\
$^{4}$ Excellence Cluster `Universe', Boltzmannstr. 2, 85748 Garching, Germany.}}

\begin{document}

\date{Accepted ???. Received ???; in original form ???}

\pagerange{\pageref{firstpage}--\pageref{lastpage}} \pubyear{2012}

\maketitle
\label{firstpage}

\begin{abstract}
Collisions between giant molecular clouds are a potential mechanism for triggering the formation of massive stars, or even super star clusters. The trouble is identifying this process observationally and distinguishing it from other mechanisms. We 
produce synthetic position--velocity diagrams from models of: cloud--cloud collisions, non-interacting clouds along the line of sight, clouds with internal radiative feedback and a more complex cloud evolving in a galactic disc, to try and identify unique signatures of collision. We find that a broad bridge feature connecting two intensity peaks, spatially correlated but separated in velocity, is a signature of a high velocity cloud--cloud collision. {We show that the broad bridge feature is resilient to the effects of radiative feedback, at least to around 2.5\,Myr after the formation of the first massive (ionising) star}. However for a head on 10\,km/s collision we find that this will only be observable from 20-30 per cent of viewing angles.  Such broad--bridge features have been identified towards M20, a very young region of massive star formation that was concluded to be a site of cloud--cloud collision by \cite{2011ApJ...738...46T}, and also towards star formation in the outer Milky Way by \cite{2014ApJ...795...66I}.


\end{abstract}


\section{Introduction}

{Understanding the galactic environment is an essential prerequisite to understanding the Galaxy's star formation rate \citep{Fujimoto2014, Hughes2013}}. Interactions between star-forming gas clouds are a major factor in cloud evolution, with events varying from tidal encounters to head-on collisions. While many of these are expected to be minor, leading to  aggregation of material \citep{2015MNRAS.446.3608D}, more major collisions are being looked to as a possible formation mechanisms for super star clusters (SSCs) and massive star formation \citep{Furukawa2009, Ohama2010, 2014ApJ...780...36F}. During such an encounter, shock compression at the collision interface leads to the rapid generation of unusually high densities compared to that found in an isolated, turbulent star-forming cloud. This high density slab can form massive cores either via an elevated Jeans mass or subsequent high accretion rates. {Such a scenario tackles the traditional problem with forming massive stars, from which} strong radiation output during formation may terminate accretion before a high mass is reached \citep{McKee2002}. However, confirming these theories has been difficult due to the challenges of identifying a collision event. Specifically:
\begin{enumerate}
\item The collision itself will happen on relatively short timescales of typically a few Myr (e.g. it takes $\sim$2\,Myr to cross a 7\,pc cloud at 10\,km/s in the absence of braking). 
\item The number of massive star-forming regions in the Milky Way is low, implying that the collision frequency is similarly small. Observational and theoretical considerations suggest collisions between giant molecular clouds (GMCs) may happen once every 10\,Myr, but the fraction of these that form stars will likely depend on morphology and collision speed \citep{Tasker2009, 2014ApJ...792...63T, fukui2015}. 
\item Massive star formation will result in feedback that can disrupt the cloud and the signature of a collision. {Even if the signature is only partially disrupted, the system may just be studied as a massive star forming region, overlooking the possibility of a collision event}. 
\item The complexity of the multiphase interstellar medium (ISM) makes it difficult to differentiate between star-forming collision events and less extreme interactions. The situation is made worse given that most collisions are expected to occur in the densely populated spiral arms of the galactic disc \citep{2013arXiv1312.3223D, Fujimoto2014}.
\end{enumerate}

Recent work by e.g. \citet[][]{2009ApJ...696L.115F, 2011ApJ...738...46T} and \citet{2014ApJ...780...36F} has highlighted the potential importance of point (iii). They identify two distinct cloud components at different velocities surrounding SSCs. While relative velocities of the two clouds remove the possibility of a single bound system, their thermal properties suggest they are interacting with the star clusters. This raises the possibility that a collision between subregions of the two clouds may have been the trigger for the formation of the SSC. {The SSCs themselves are observed to reside at the junction between the clouds, further supporting this idea}. Recent numerical simulations by \citet{2015MNRAS.446.3608D} support the idea that collisions do occur over only small subsets of the clouds, leaving the extended gas to retain the kinematic structure of the pre-collision cloud, in qualitative agreement with the aforementioned observations. 

Collisions between GMCs have been studied using hydrodynamical models both in smaller-scale dedicated collision simulations \citep[e.g.][]{2013ApJ...774L..31I, 2014ApJ...792...63T} and in the context of the entire galaxy \citep[e.g.][]{2006MNRAS.367..873D, 2011MNRAS.417.1318D, Tasker2009, 2011ApJ...730...11T, 2015MNRAS.446.3608D, Fujimoto2014}. While only the former are capable of achieving the resolution required to follow the collapse into cores, the global simulations provide a clue to the rate at which such interactions occur. They agree that interactions for a cloud should occur multiple times per orbital period, a rate that could potentially drive the entire star formation rate of the galaxy if all collisions were productive \citep{Tan2000, Fujimoto2014b}. The local-scale simulations and observations, however, suggest collision results are velocity dependent.

Recent N--body simulations by \cite{2014ApJ...787..158B, 2015MNRAS.447..732B} found that the massive star cluster NGC 3603 favours a monolithic formation mechanism over hierarchical assembly, the former of which is consistent with widespread massive star formation triggered by collisions (although not exclusively).

Synthetic observations based on simple cloud collision models have also been tried. \cite{1989ApJ...346..184K} modelled collisions between uniform density high latitude clouds and then post-processed their results to produce optically thin $^{13}$CO emission maps and line profiles. They found these could explain many of the observed features in high latitude clouds, including secondary peaks and broad wings. More recently, \cite{2011A&A...528A..50D} created  surface density position-velocity diagrams from hydrodynamical models of colliding cylinders (which, seeded with a turbulent field {quickly evolved into a collection of filamentary structures}). They found that their results provide a possible explanation for the trigger of recent star formation in Serpens. The clouds in this latter work were smaller than GMCs (the cylinders had initial radii of 0.25\,pc and a typical length of 1\,pc). 

In this paper we post-process a set of hydrodynamic and radiation hydrodynamic simulations, covering a range of possible scenarios including: non-interacting clouds coincident along the line of sight, a cloud that has evolved in a galactic environment, cloud--cloud collisions of different velocities and clouds with internal radiative feedback. 
Using these models we  produce synthetic $^{12}$CO position--velocity (p--v) diagrams which we use search for any signatures that are characteristic to a major cloud--cloud collision of the kind that might give rise to massive star formation. If one can determine the expected observational characteristics of enough theoretical models, any non-degenerate signatures should be very useful for interpreting real observations. 


\begin{figure}
	\vspace{-10pt}
	\includegraphics[width=7.5cm, height=7.5cm]{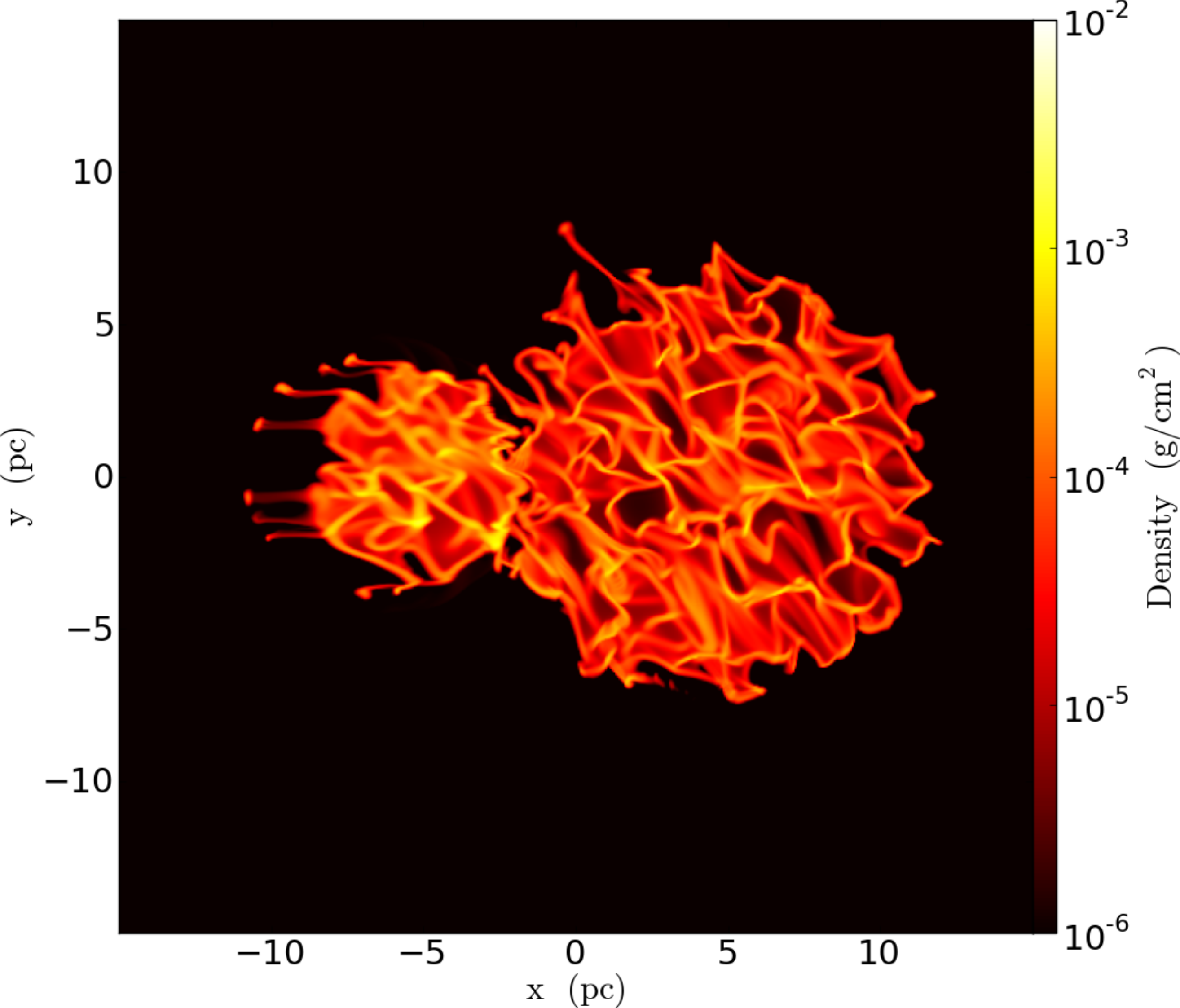}
	\includegraphics[width=7.5cm, height=7.5cm]{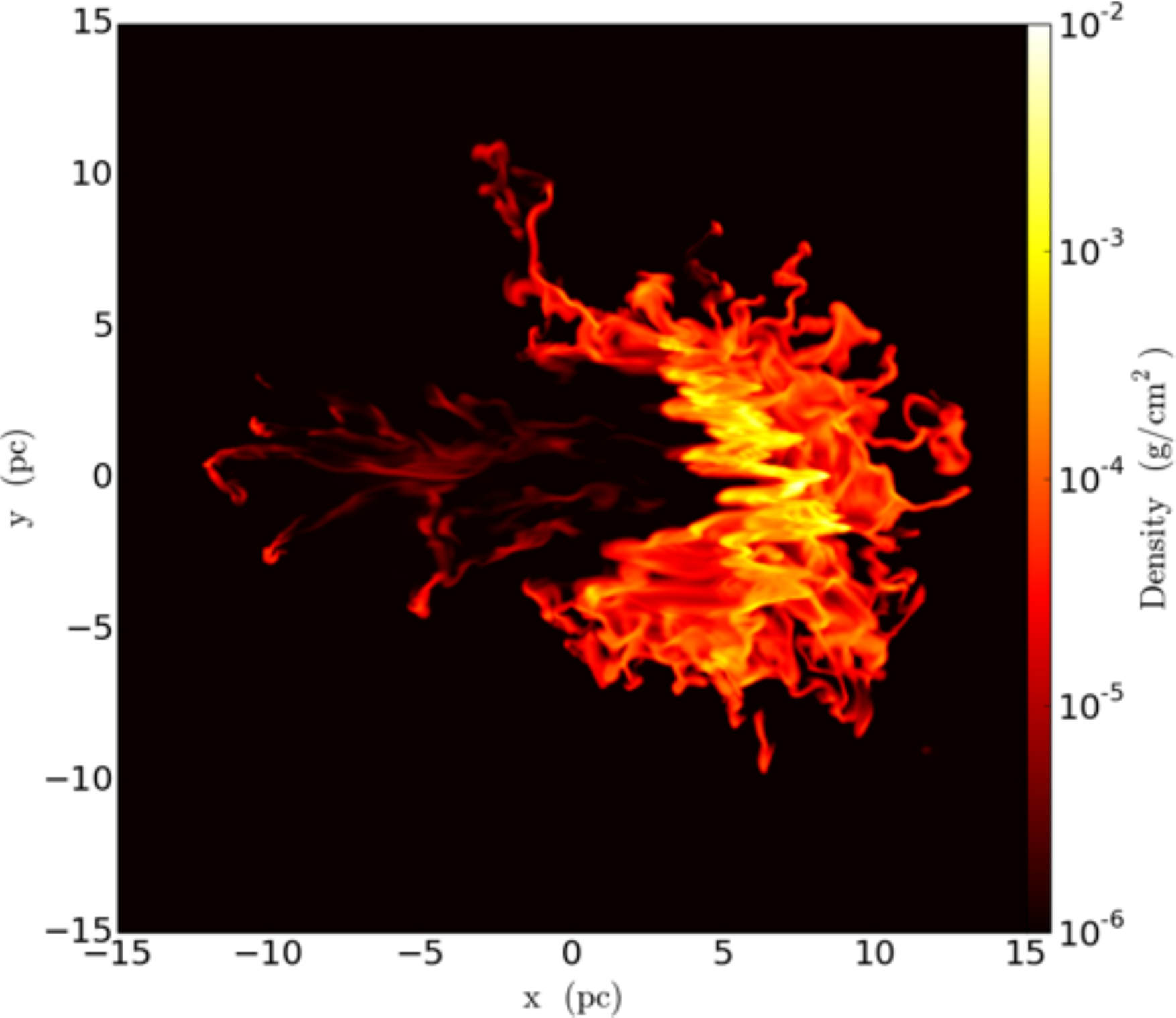}	
	\includegraphics[width=7.5cm, height=7.5cm]{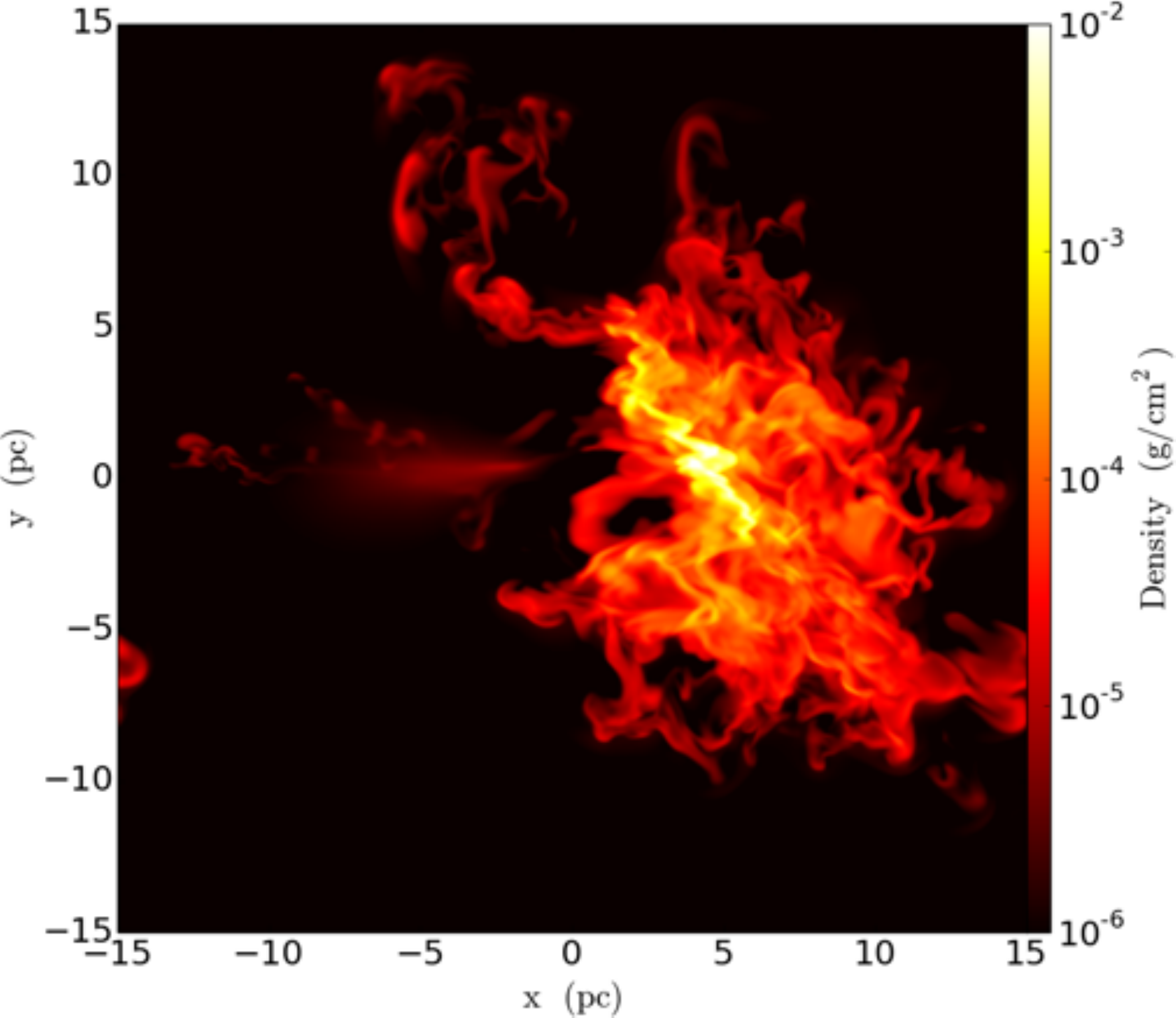}
	\caption{Surface density plots of the \textsc{enzo} cloud--cloud collision simulation snapshots. The upper panel shows the clouds just prior to a 10\,km/s collision. The middle and lower panels show the 10 and 3\,km/s collision models respectively. All snapshots are at a viewing angle perpendicular to the collision axis.  }
	\label{enzohydro}
\end{figure}

\section{Numerical methods}
\subsection{Hydrodynamic models}

\subsubsection{Hydrodynamics simulations with ENZO}
\label{enzo}
Two of our hydrodynamic models of cloud--cloud collisions come from runs performed in \cite{2014ApJ...792...63T}, using the grid-based hydrodynamics code, \textsc{enzo} \citep{Enzo}. The runs consist of two non-identical, turbulent, idealised clouds in a head-on collision. {In one simulation the collision velocity is 3\,km/s and in the other it is 10\,km/s}. The large and small clouds involved in the collision have radii of 7.2 and 3.5\,pc, initial turbulent velocity dispersions of 1.71 and 1.25\,km/s and mean number densities of 25.3 and 47.4\,cm$^{-3}$, respectively (i.e. the clouds have masses 417 and 1635\,M$_\odot$ respectively). The clouds are initially spherical and seeded with a turbulent velocity field. The collision itself is head on, in a frame such that the larger cloud has no bulk velocity with the smaller cloud moving towards it. We consider a snapshot of each simulation at the point where the maximum number of star-forming gas cores have formed, with a core being defined as gas within a contour of density of $0.3\times 10^4$\,cm$^{-3}$. The surface density of the 3 and 10\,km/s collision snapshots is given in the middle and lower panels of Figure~\ref{enzohydro} respectively. Additionally, we consider one extra snapshot from the 10\,km/s collision model just prior to cloud contact, to isolate the signature of two clouds coincident along the line of sight, but not interacting. We include this pre-collision snapshot in the upper panel of Figure~\ref{enzohydro}.  The limiting resolution (smallest cell size) in these simulations was 0.06\,pc.

\begin{figure}
	\includegraphics[width=8cm, height=8cm]{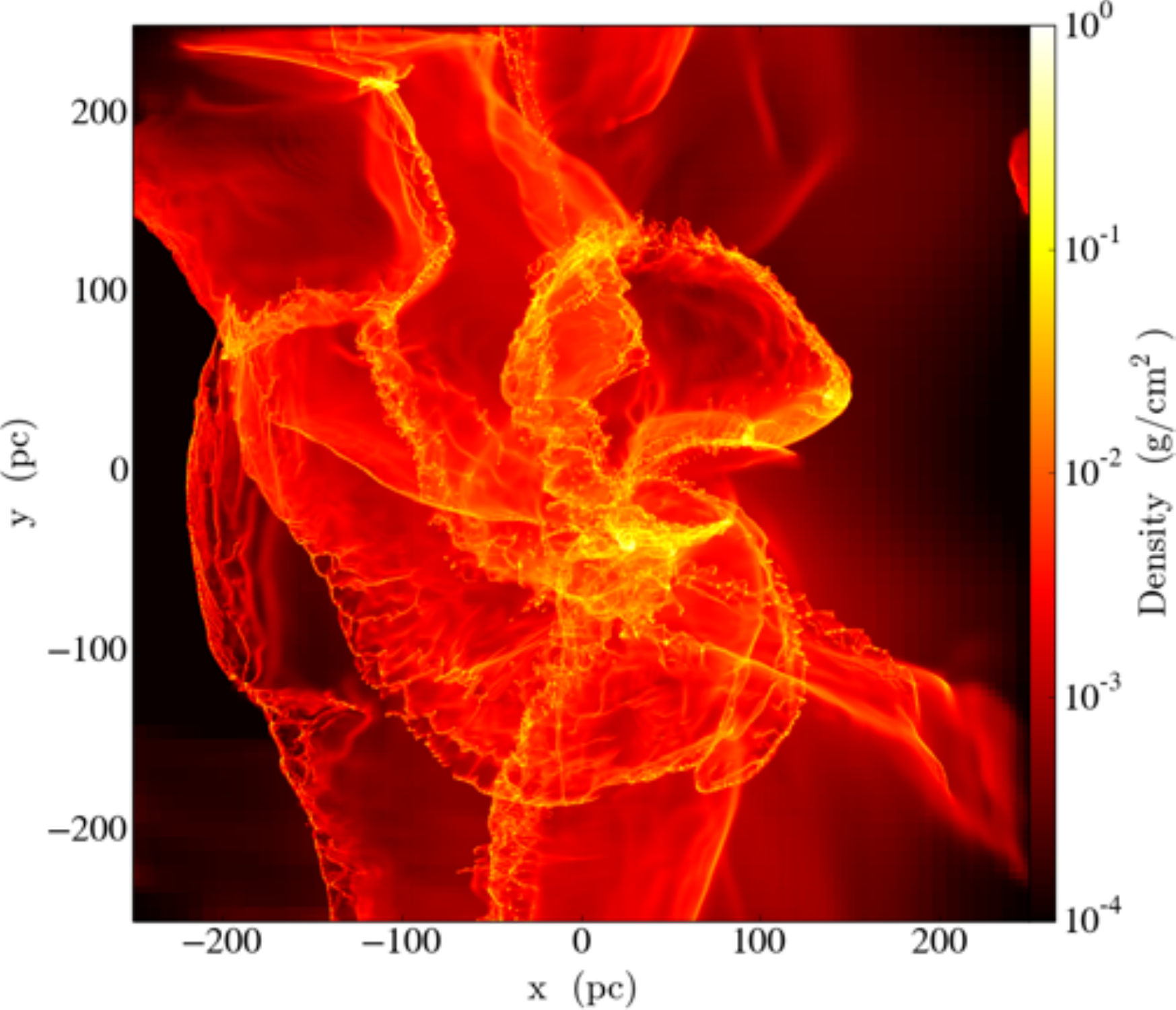}
	\caption{A surface density snapshot of the cloud extracted from a galactic scale \textsc{enzo} simulation.  }
	\label{enzogal}
\end{figure}

A third hydrodynamic model involved in this comparison was of a star-forming cloud formed in a galaxy-scale simulation from  \cite{2013ApJ...776...23B}. The cloud was extracted from the galaxy disc and evolved at higher resolution (Shima, Tasker and Habe, in prep.) to give a smallest cell size of 0.1\,pc. The surface density distribution of this cloud is shown in  Figure~\ref{enzogal}.

\subsubsection{Radiation hydrodynamic feedback models with SPH-NG}
\begin{figure}
	\includegraphics[width=7.5cm, height=7.5cm]{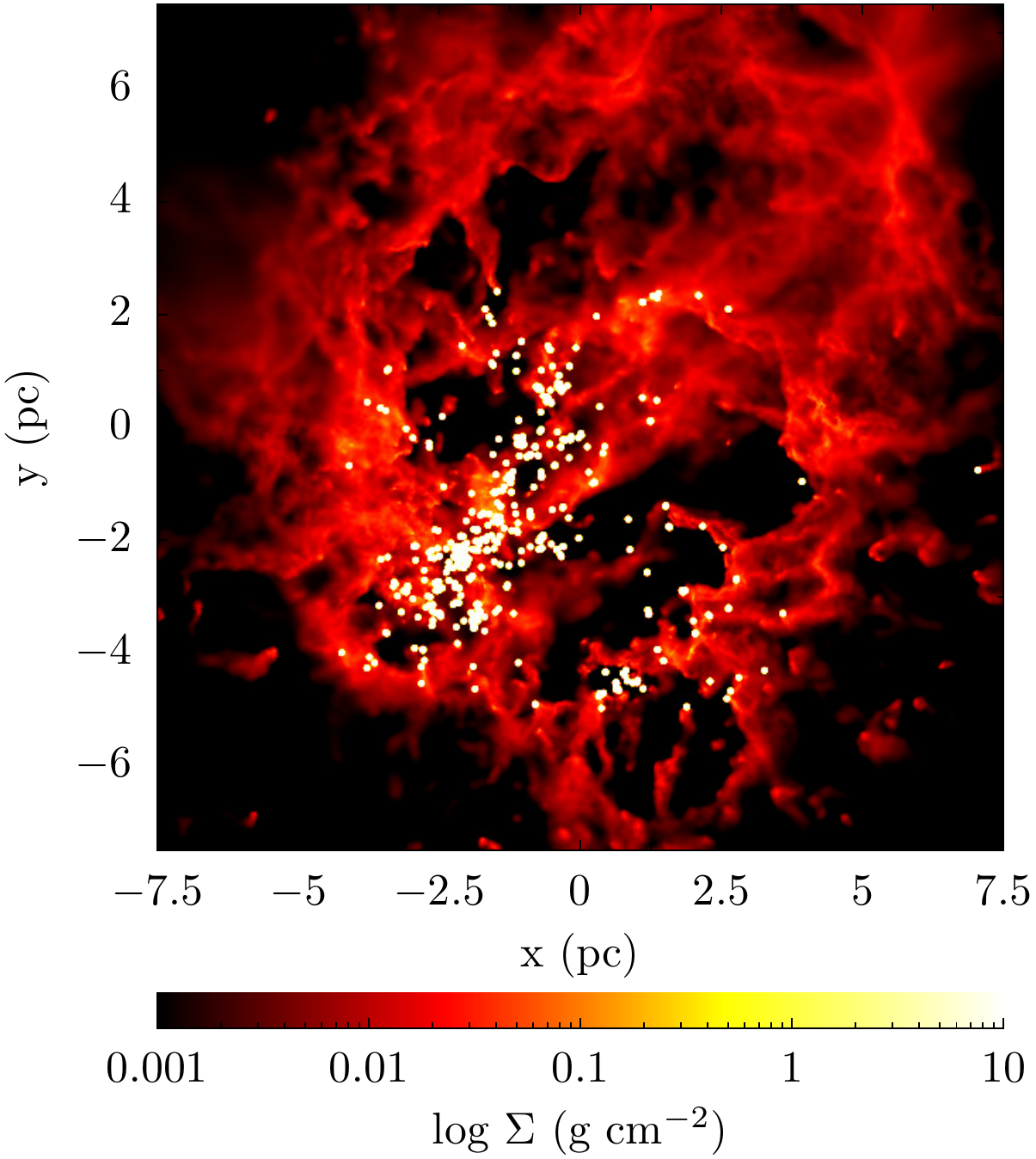}
	\includegraphics[width=7.5cm, height=7.5cm]{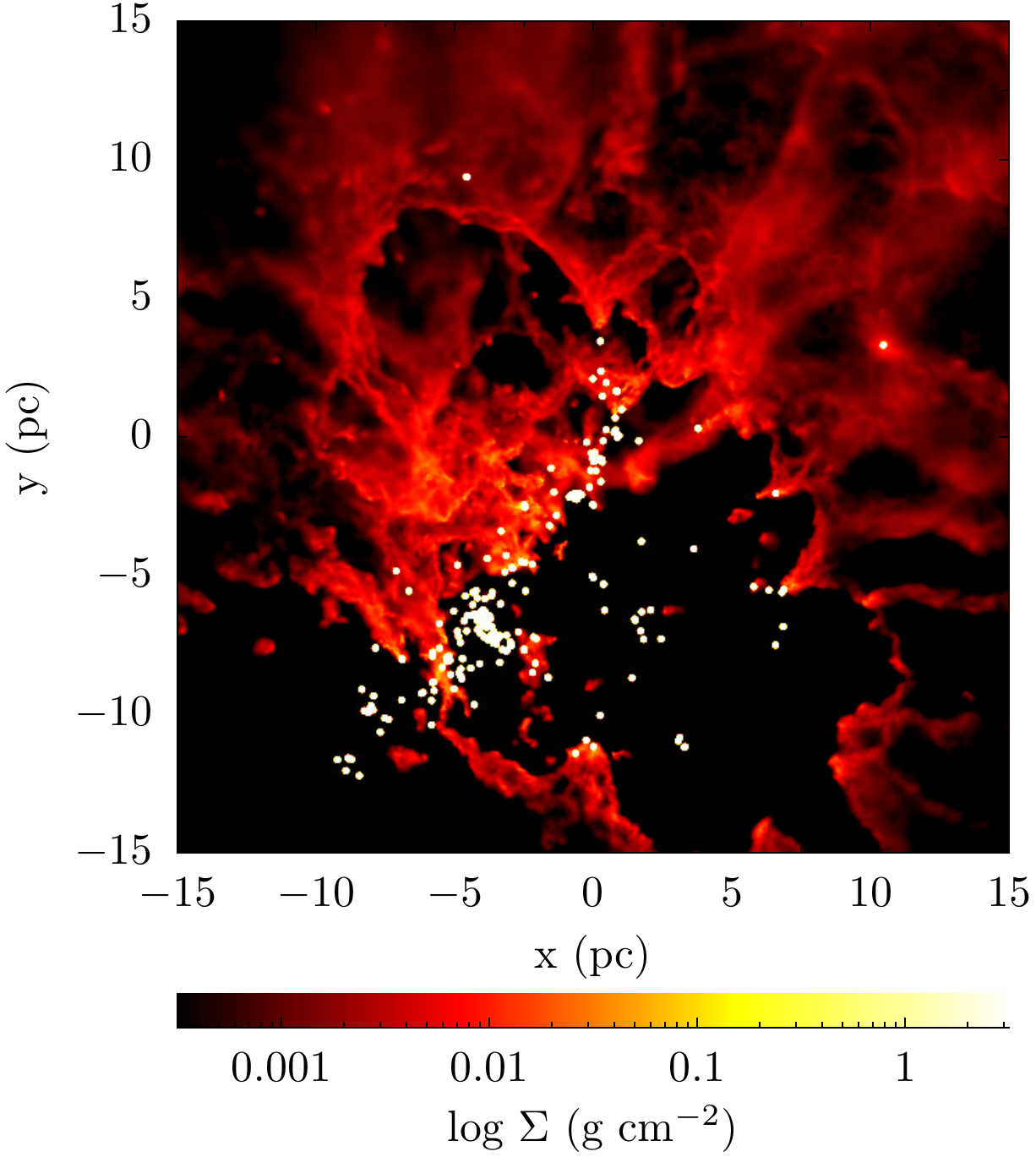}	
	\includegraphics[width=7.5cm, height=7.5cm]{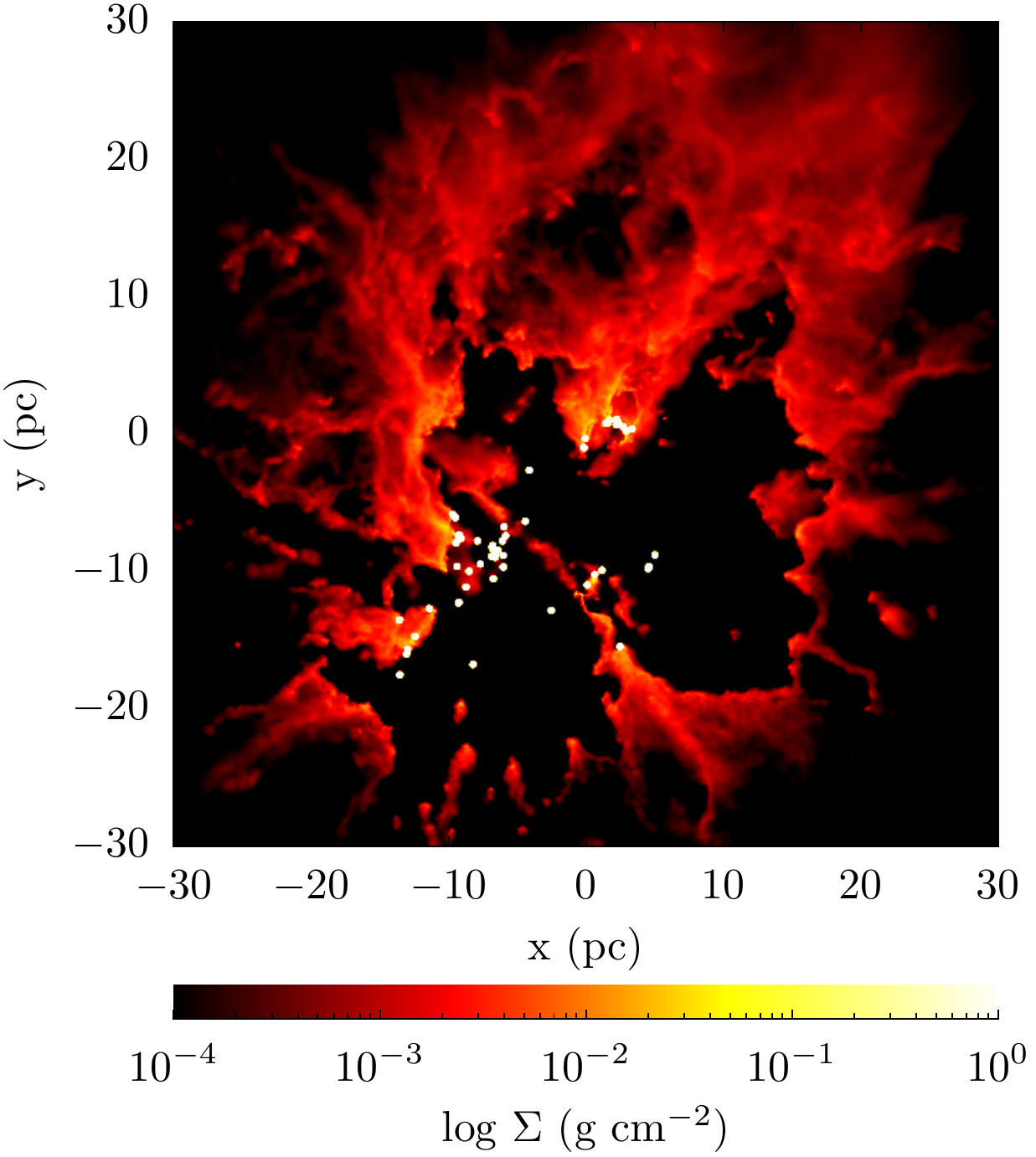}		

	\caption{Surface density plots of the \textsc{sph-ng} simulation snapshots used in this paper. From top to bottom the panels are models J, UP and UQ from Dale et al. White dots are sink particles.  }
	\label{hydro}
\end{figure}

To distinguish between observational signatures due to collisions and those due to stellar radiative feedback we also use 3 of the radiation hydrodynamic simulations of massive star feedback discussed in a series of papers by Dale et al., e.g. \cite{2012MNRAS.424..377D} and  \cite{2013MNRAS.431.1062D}, namely models J, UP and UQ. These are smoothed particle hydrodynamics simulations of radiative feedback from massive stars in turbulent clouds. They followed the gravitational collapse of a turbulent cloud and incorporated ionisation effects from any stars (sink particles) that attain mass $\geq20$\,M$_\odot$. The effect of the ionising radiation is to heat the gas around the stars, resulting in hot ionised bubbles partially surrounded by molecular gas that is accelerated away from the ionising sources. Surface density plots of the snapshots we use are given in Figure \ref{hydro}. In model J the initial cloud is gravitationally bound, whereas in UP and UQ they are initially partially gravitationally unbound.

\subsection{Radiative transfer}
We use the \textsc{torus} radiation transport and hydrodynamics code to produce synthetic observations in this paper \citep{2000MNRAS.315..722H,2010MNRAS.407..986R, 2012MNRAS.420..562H}. Specifically we use the non-LTE molecular line transfer developed by and discussed in great detail in \cite{2010MNRAS.407..986R} and summarised in \cite{2013MNRAS.431.3470H}. Given that it is documented comprehensively in these papers, we do not discuss the algorithm in any detail here. 

We map the results (densities, temperatures, velocities) of the various simulations onto the \textsc{torus} grid. The SPH data is mapped onto the \textsc{torus} grid using the method described in \cite{2010MNRAS.403.1143A}. We assume a fixed molecular abundance ($8\times10^{-5}$ relative to hydrogen for $^{12}$CO) and solve for the molecular level populations using a non-LTE statistical equilibrium calculation. We do include dust in these calculations and consider a standard ISM power law size distribution \citep{1977ApJ...217..425M} with optical coefficients from \cite{1984ApJ...285...89D} and a dust to gas ratio of $10^{-2}$. Once the level populations are converged {they are used in further ray tracing calculations to produce synthetic position--position--velocity data cubes for as many observer viewing angles as desired \citep{2010MNRAS.407..986R}. We use the \textsc{starlink} software \textsc{gaia} to collapse these data cubes along one spatial axis to transform them into p--v diagrams. }

We do not convolve the resulting p--v diagram to a beam representative of any real instrument since we are not trying to replicate a specific set of observations in this paper. The spectral and spatial resolution in the cloud--cloud collision models are 0.04\,km/s and 0.2\,pc.

\subsection{Summary of numerical models}
We post process a set of 7 numerical models to generate synthetic $^{12}$CO J=1-0 p--v diagrams for clouds coincident along the line of sight, clouds undergoing collision,  clouds subject to internal radiative feedback and a cloud that has evolved in a galaxy-like environment. For ease of reference and navigation of the paper we summarise the models in Table \ref{models}. Note that in section \ref{limits} we will also briefly consider the case of a collision between clouds, including internal radiative feedback.

\begin{table*}
 \centering
  \caption{Summary of the numerical models post processed in this paper.}
  \label{models}
  \begin{tabular}{@{}l l c c c@{}}
  \hline
   Model & Scenario & Origin & Surface densiity & p--v diagram\\
   ID & & & snapshot \\
  \hline   
   A & Turbulent clouds coincident along the line of sight & Takahira et al (2014) & Fig 1, top panel &Fig 5, top panel\\  
   B & Turbulent clouds collided at 10\,km/s & Takahira et al (2014) & Fig 1, middle panel & Fig 5, middle panel\\  
   C & Turbulent clouds collided at 3\,km/s & Takahira et al (2014) & Fig 1, bottom panel & Fig 5, botom panel\\  
   D & A cloud that has evolved in a galactic disc & Benincasa et al. (2013) & Fig 2 & Fig 8\\     
	& & (Shima et al. in prep.) & & \\
   J & A  gravitationally bound cloud with  & Dale et al (2012, 2013) & Fig 3, top panel & Fig 6, top panel\\  
	& internal radiative feedback & & &\\
   UP & A  partially gravitationally unbound cloud with & Dale et al (2012, 2013)  & Fig 3, middle panel & Fig 6, middle panel\\  
   &  internal radiative feedback & &  &\\
   UQ & A partially gravitationally unbound cloud with & Dale et al (2012, 2013)  & Fig 3, bottom panel & Fig 6, bottom panel\\  
   &  internal radiative feedback & &  &\\
\hline
\end{tabular}
\end{table*}

\section{Results of synthetic observations}

\begin{figure}
	\hspace{0pt}	
	\includegraphics[width=8cm]{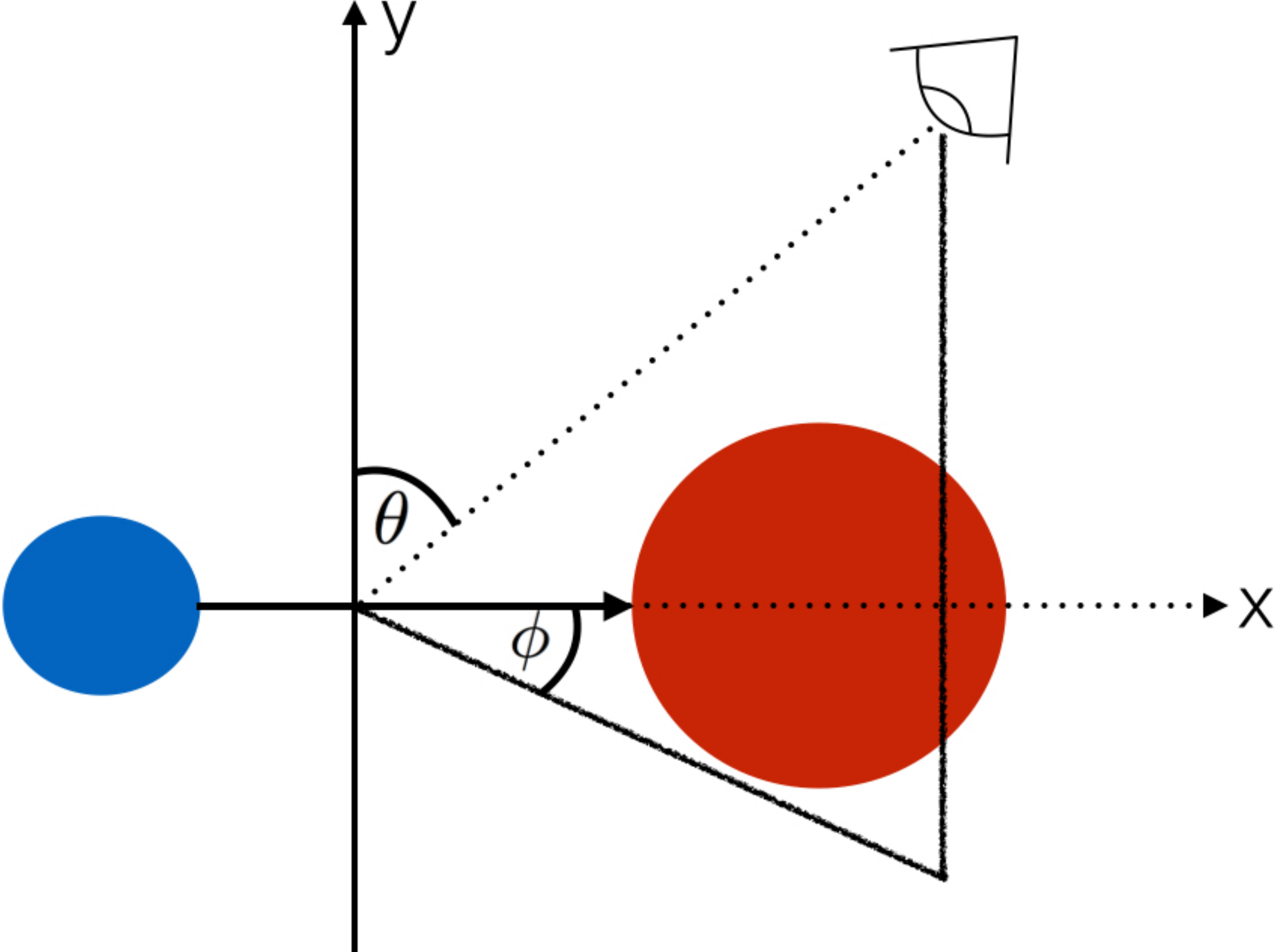}	
	\caption{A schematic of the viewing angle convention used in this paper. A viewing angle of $\theta=\rm{\pi}/2$, $\phi=0$ is along the collision axis. {Note that the collision simulations take place in a frame in which the larger cloud is stationary.} }
	\label{schematic}
\end{figure}

\subsection{p--v diagram morphology}

We calculated $^{12}$CO J=1-0 p--v diagrams for all of the numerical models summarised in Table \ref{models}. They are constructed by integrating along the entire $x$-axis at each velocity as a function of the $y-$axis on the image grid.  The results are presented in Figures 5 through 8. We now discuss the different p--v diagram morphologies.

\subsubsection{Collisional models}

\begin{figure}
	\includegraphics[width=7cm, height=7cm]{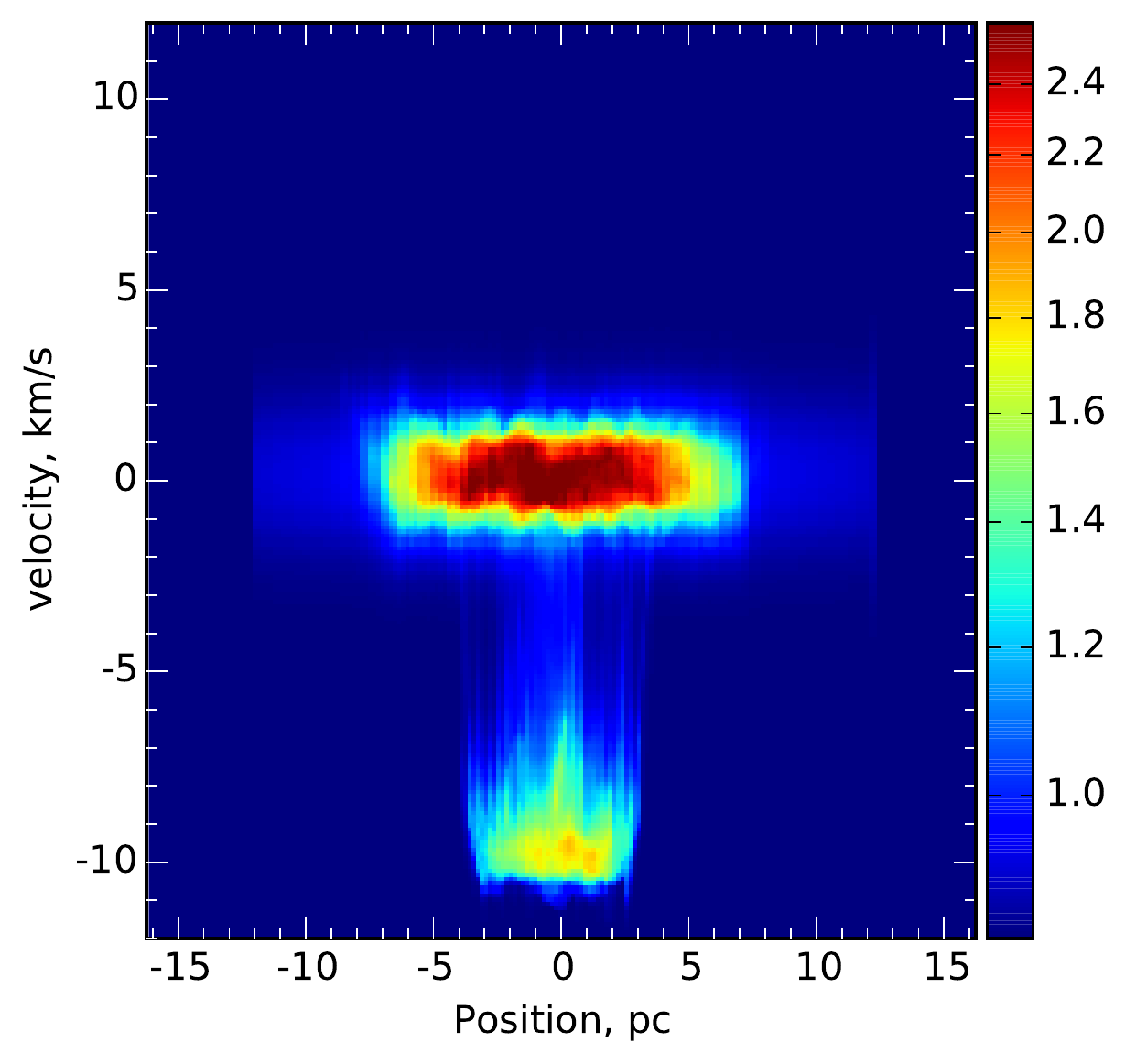}	
	\includegraphics[width=7cm, height=7cm]{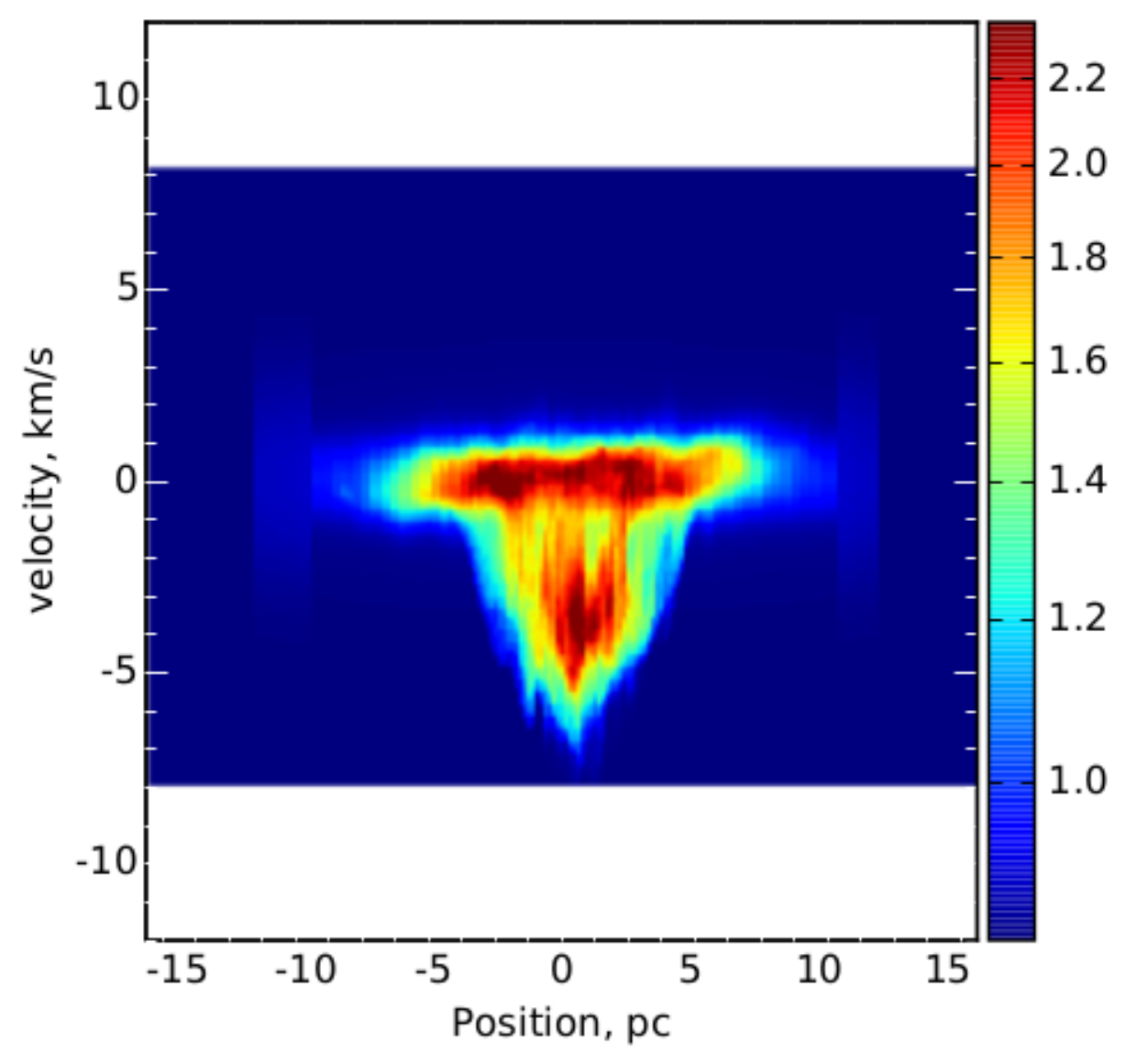}
	\includegraphics[width=7cm, height=7cm]{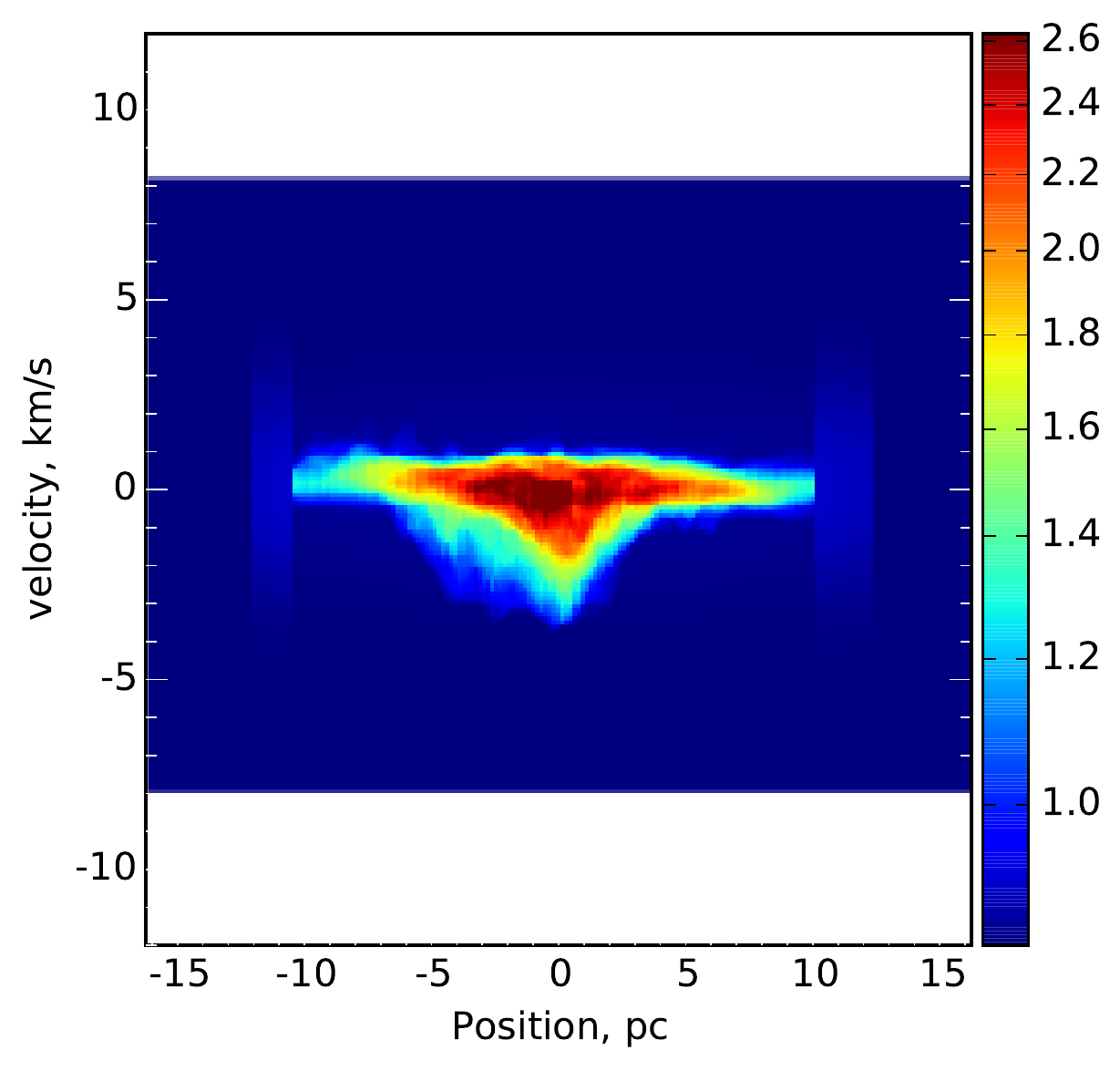}		
	\caption{$^{12}$CO  J=1-0 position--velocity diagrams of the \textsc{enzo} models of cloud--cloud collisions. The upper panel is for a simulation snapshot just prior to the collision (the clouds are just touching) in the 10\,km/s collision model, with both clouds along the line of sight. The middle and lower panels are from the 10 and 5\,km/s collision models respectively at the point of maximum core formation in each model (c.f. \ref{enzo}).}
	\label{pvmaps_col}
\end{figure}

The p--v diagrams resulting from the collisional models (models A-C in Table \ref{models}) are given in Figure \ref{pvmaps_col}. The upper panel shows the p--v diagram for the two clouds in the 10\,km/s collision model at a time just prior to the collision (model A) viewed along the axis of collision such that the clouds are coincident along the line of sight ($\theta$=$\pi/2$, $\phi$=0 as defined in Figure \ref{schematic}). Both clouds are clearly separated in velocity space by the known pre-collision velocity of 10\,km/s. Each individual cloud has a width in velocity space determined by their turbulent velocity dispersion. The low intensity, intermediate velocity gas is from material at the interface between the small and large clouds that has begun to collide/brake (the clouds are just touching at this snapshot in time). 

The middle and lower panels of Figure \ref{pvmaps_col} show the 10 and 3\,km/s collision p--v diagrams respectively, also viewed along the collision axis ($\theta$=$\pi/2$, $\phi$=0). Compared to the merely coincident clouds these diagrams are rather different. In the 10\,km/s collision case, the two peak features are separated by lower velocity due to braking and are connected by a broad bridge feature. This signature is composed of the remnants of the two clouds (giving rise to the two peaks) and intermediate velocity gas at the interface between the two clouds (giving rise to the bridge). 

In the 3\,km/s case two peaks are not discernible and there is therefore not any broad bridge feature. This is because the clouds initially have a smaller velocity difference compared to the turbulent velocity dispersion of the clouds, making the bridge harder to distinguish, and this difference is rapidly reduced during the collision due to braking. The only signature different from that of an isolated turbulent cloud is the spike in negative velocity at $\sim0$\,pc, which comes from shocked gas. If the local turbulent velocity is lower it is possible that lower velocity collisions might be identified, for example \cite{2011A&A...528A..50D}  have identified double velocity components separated by bridges in simulations of elongated collisions between low mass clouds at only 2\,km/s. 

It therefore seems that a broad bridge separating two peaks and evidence of shocked gas are signatures of cloud--cloud collision, though whether this is identified is sensitive to the collision velocity relative to the turbulent velocity of the colliding gas. We will explore this velocity sensitivity further in section \ref{angles}.

\begin{figure}		
	\includegraphics[width=7cm, height=7cm]{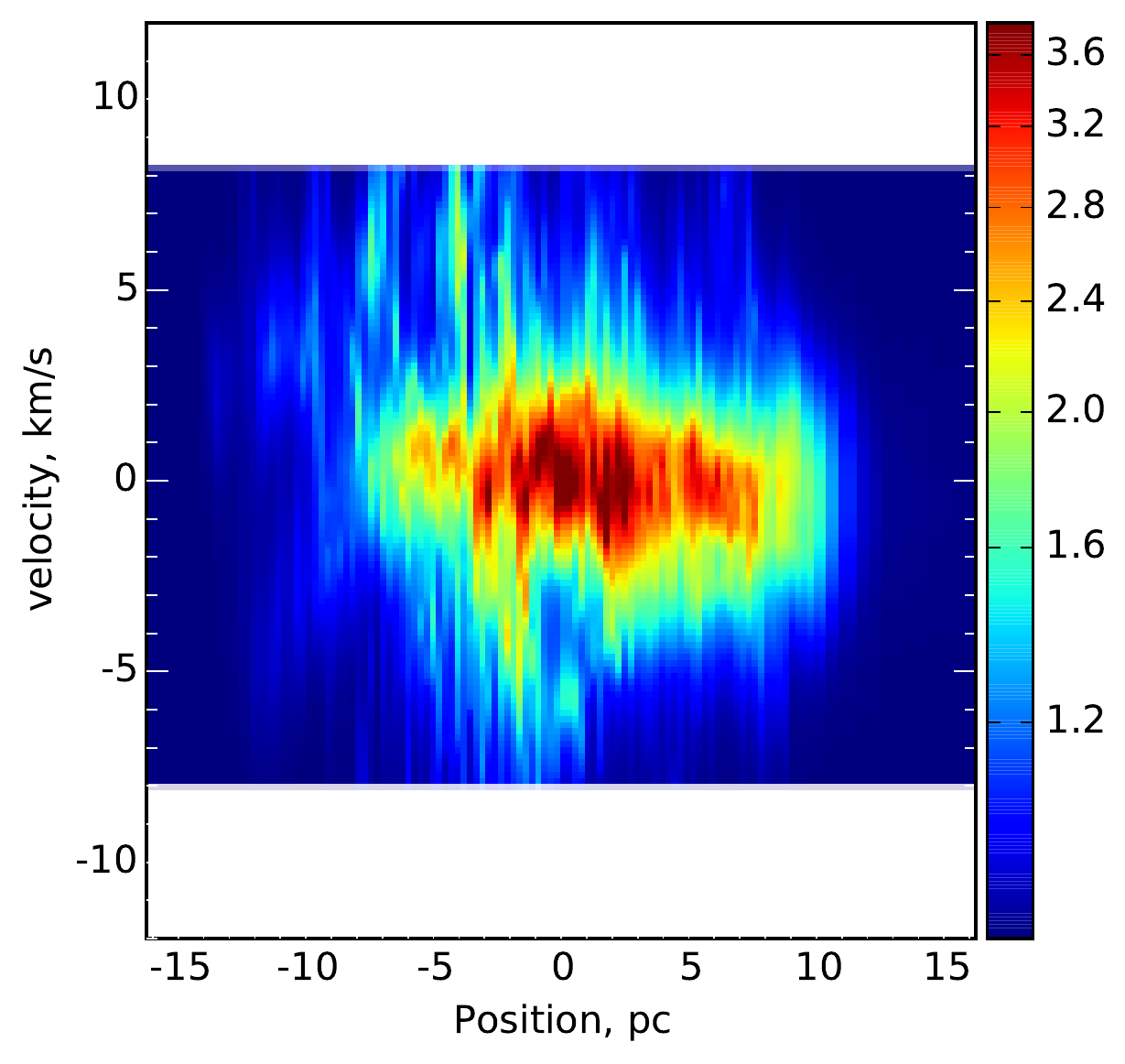}	
	\includegraphics[width=7cm, height=7cm]{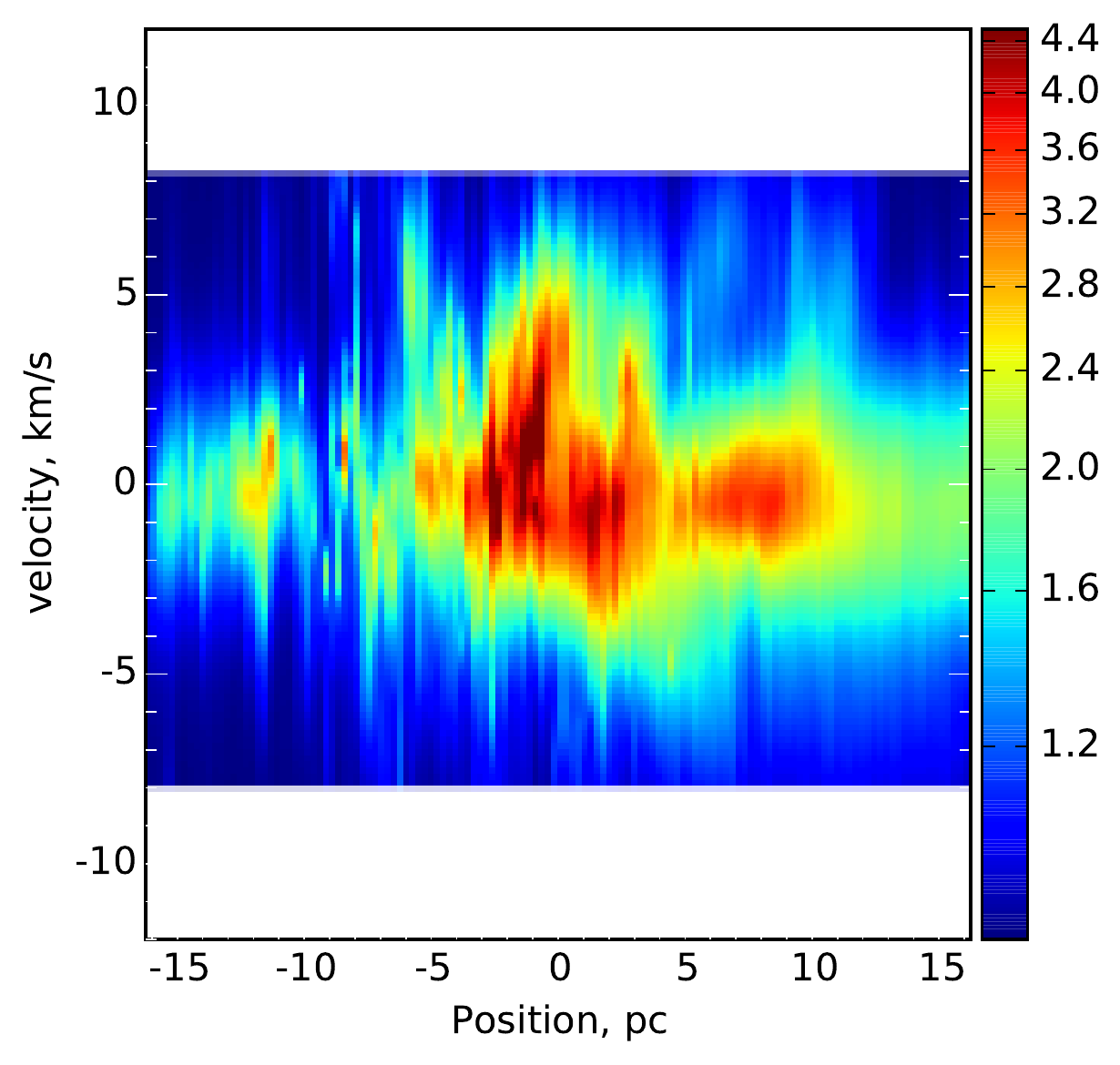}	
	\includegraphics[width=7cm, height=7cm]{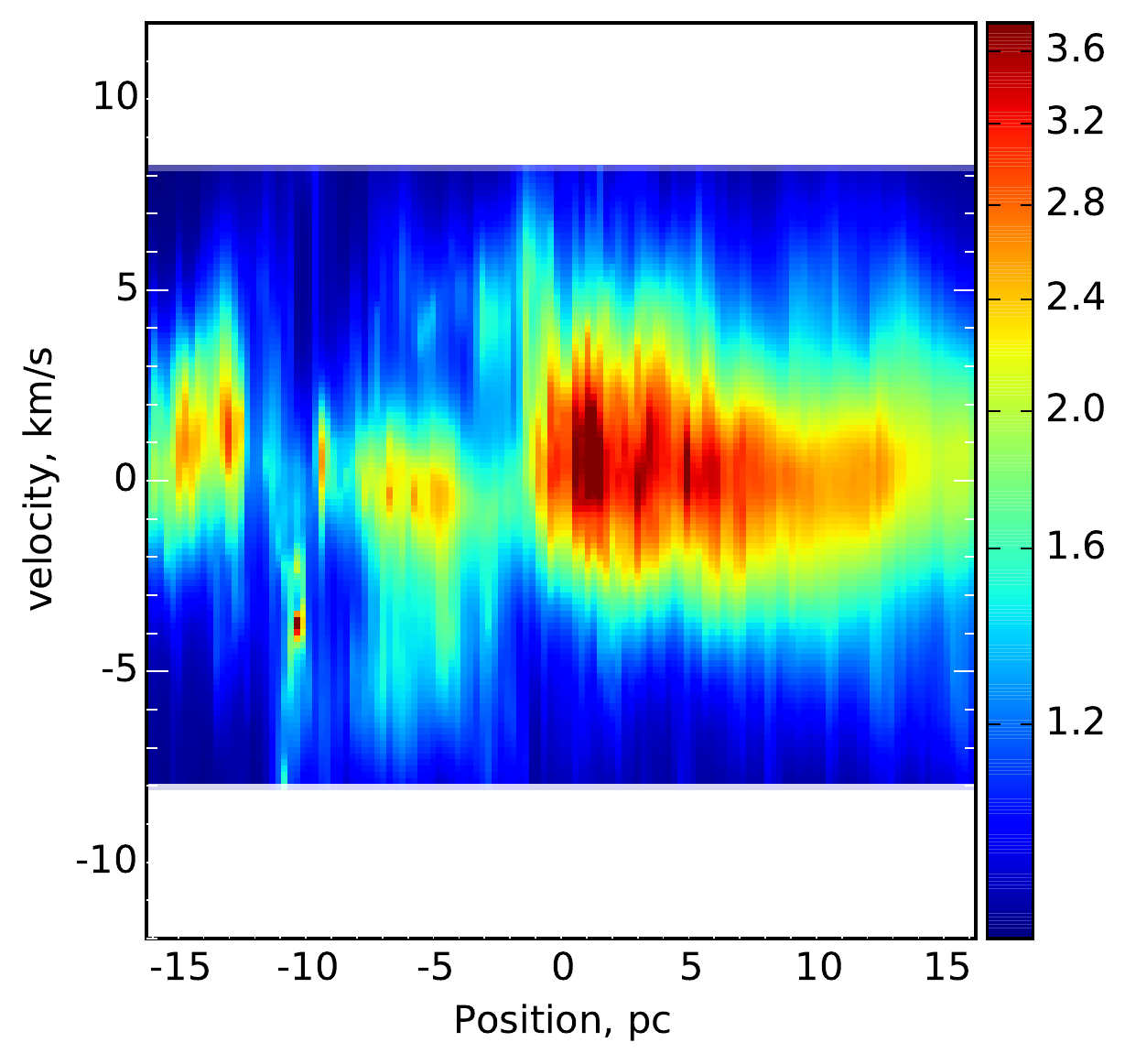}		
	\caption{$^{12}$CO J=1-0 position--velocity diagrams for the feedback simulations of Dale et al. From top to bottom the panels are models J, UP and UQ. }
	\label{pvmaps_rad}
\end{figure}

\begin{figure}
	\hspace{-20pt}
	\includegraphics[width=9cm]{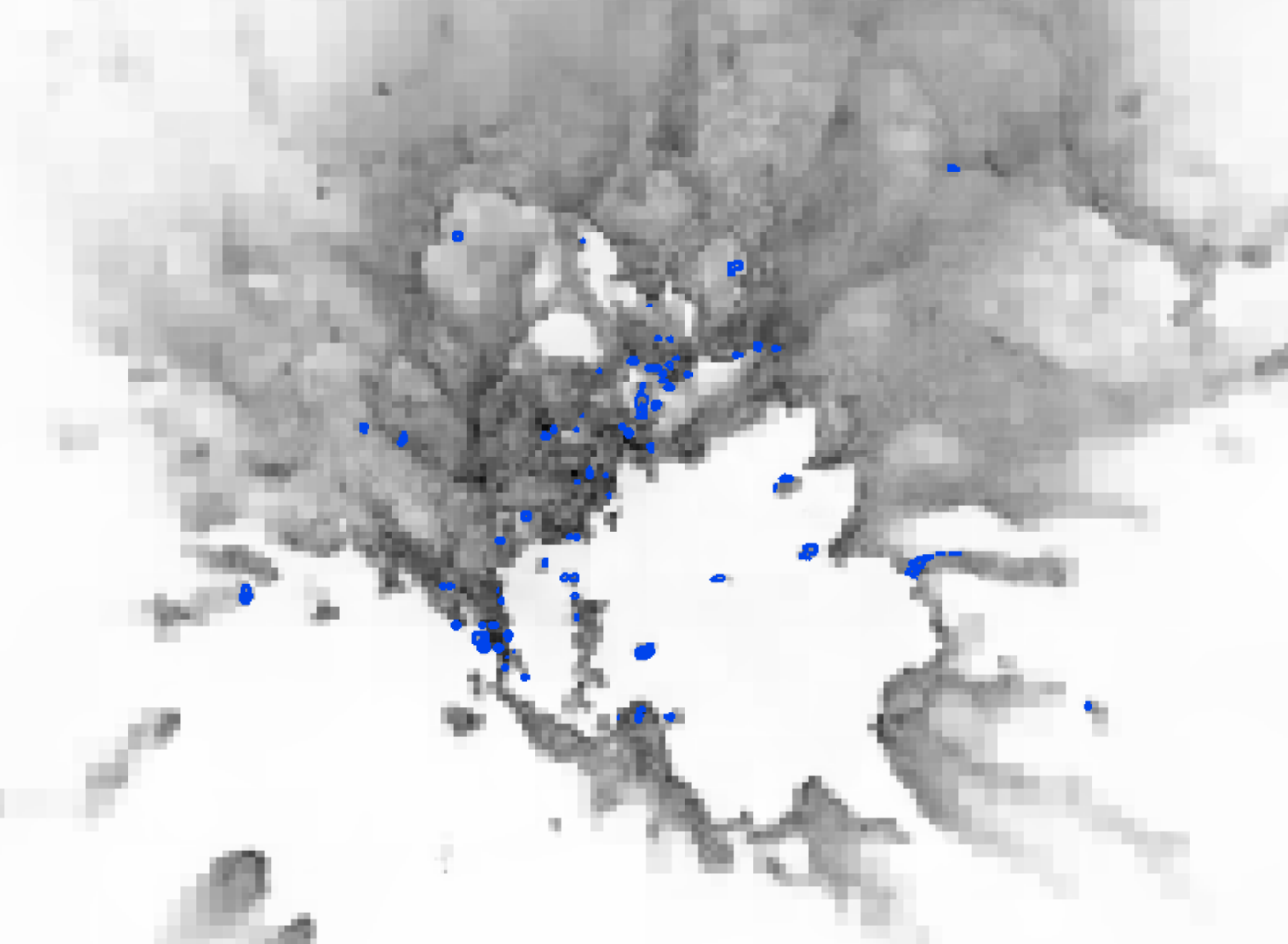}	
	\caption{Integrated CO J=1-0 emission from model UP with contours overlaid from the highest and lowest velocity channels. High velocity features are globules within the H\,\textsc{ii} region and accelerated gas at the H\,\textsc{ii} region boundary.}
	\label{highvelUP}
\end{figure}

\subsubsection{Radiative feedback models}
\label{radf}
The p--v diagrams resulting from the radiative feedback models (models J, UP and UQ in Table \ref{models}) are given in Figure \ref{pvmaps_rad}. The panels are model J, UP and UQ from top to bottom. 

In the classical picture of an expanding H\,\textsc{ii} region, where there is a spherical shell of dense gas bounding the ionised gas, one would expect a circular or elliptical signature in the p--v diagram. However in our results each diagram consists of a single broad feature in velocity space, with the partially gravitationally unbound clouds consisting of multiple components along the spatial axis. This signature arises rather than the elliptical one because these H\,\textsc{ii} regions are actually very leaky, losing up to 95 per cent of the ionising photons \citep[this has also been suggested observationally, by e.g.][]{2010ApJ...709..791B}. Only a small fraction of $4\pi$ centred on the ionising sources is subtended by dense molecular material and so only a small fraction of the classical sphere is actually accelerated. 

The p--v diagrams also show small spatial scale, high velocity features, which are globules (interior to the H\,\textsc{ii} region) or clumps of gas at the edge of the H\,\textsc{ii} region accelerated by the ionising radiation field. We illustrate this in Figure \ref{highvelUP} by showing the integrated emission of model UP with contours from the highest velocity channels overlaid. 

Overall these p--v diagrams of radiative feedback models show an imprint of the initial turbulent velocity structure coupled with relatively localised components of the cloud accelerated by the ionising radiation field. The p--v diagrams for models with radiative feedback unsurprisingly show signatures of high velocity gas, however none show the broad bridge feature that we find in our 10\,km/s model, supporting its use as a signature of cloud--cloud collision.

\subsubsection{A larger cloud from a galactic disc}
\begin{figure}	
	\hspace{10pt}		
	\includegraphics[width=7cm, height=7cm]{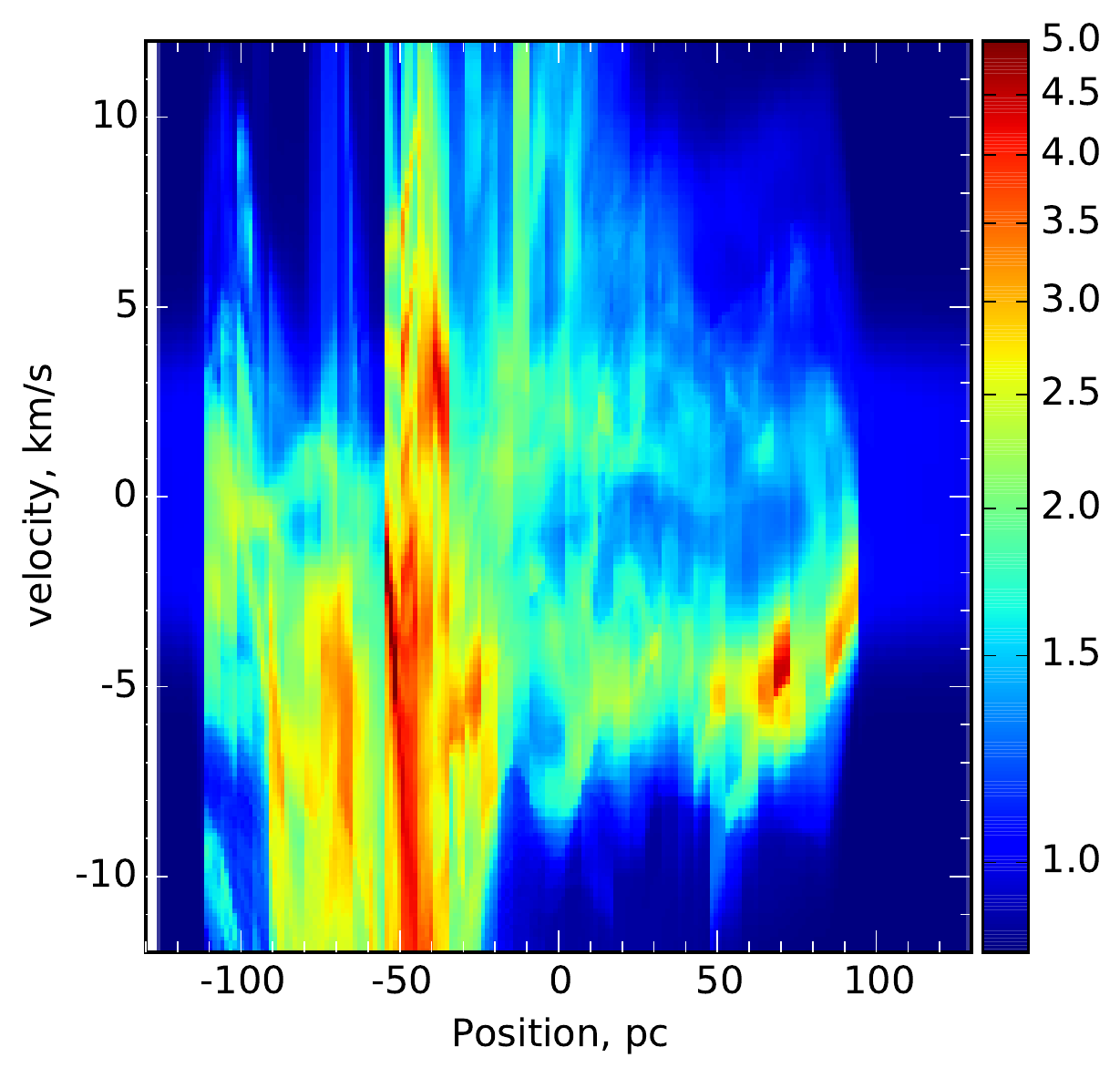}

	\hspace{10pt}			
	\includegraphics[width=7cm, height=7cm]{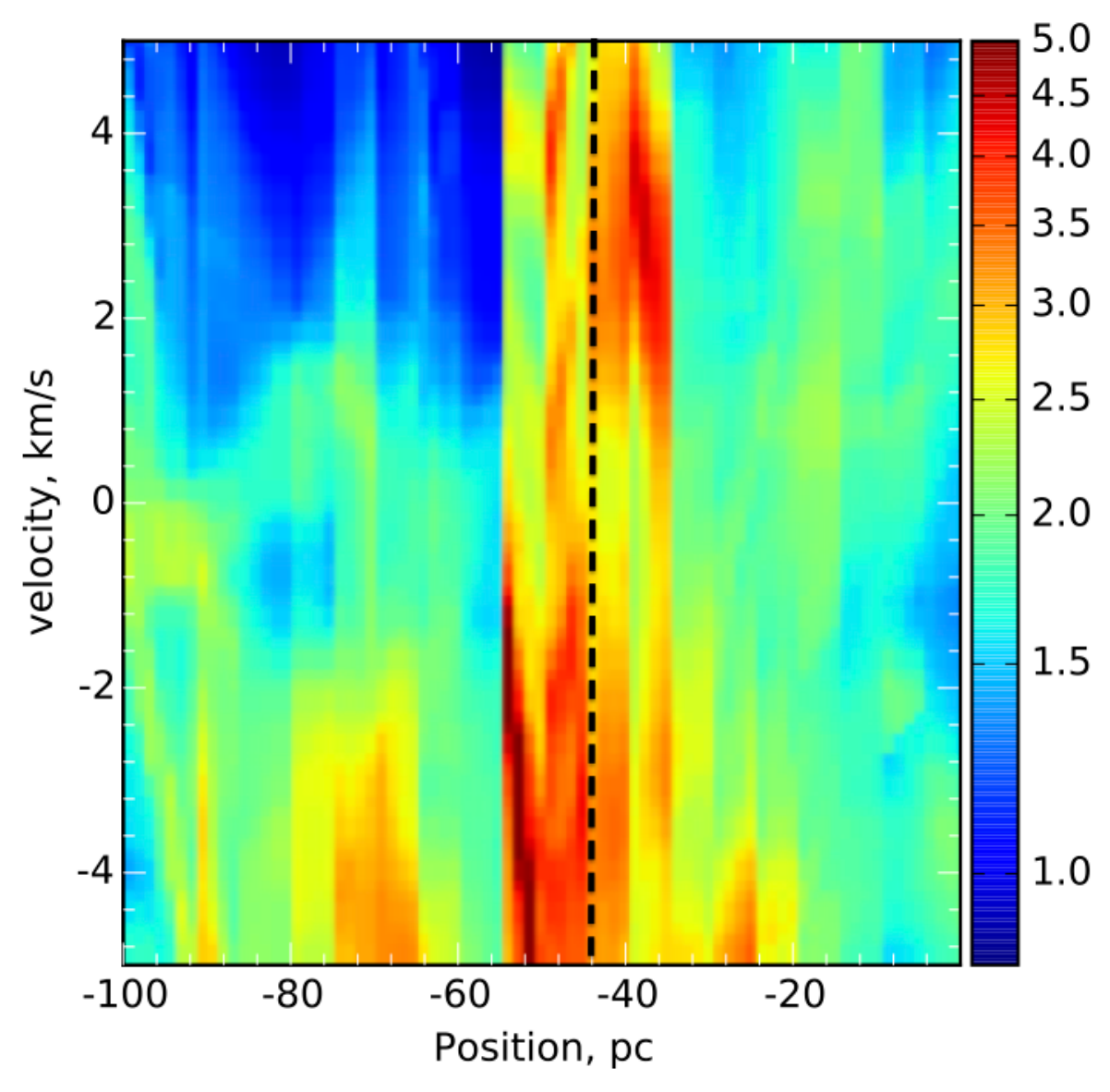}	
	\caption{$^{12}$CO J=1-0 position--velocity diagrams for the complex GMC from a galactic scale model. {The upper panel shows the whole p--v diagram and the lower zooms in on the region at $\sim$-50\,pc to illustrate that there is no broad bridge feature since the intensity peaks are spatially offset.}}
	\label{pvmaps_gal}
\end{figure}

Figure \ref{pvmaps_gal} shows the p--v diagram for the more complicated cloud taken from galaxy scale simulations of \cite{2013ApJ...776...23B}. Of course a single cloud is not really representative of the whole range of clouds evolving in a galactic disc (probing this full range is beyond the scope of this paper), however we include one to compare with a more complicated cloud structure. This cloud has evolved in a Milky Way-type galaxy without a grand design spiral where the environment is highly dynamic, it is therefore complex in geometry due to gravitationally driven mergers and tidal events happening regularly. This cloud is also much larger than the other clouds considered here.

The p--v diagram is highly disordered and  very broad along both spatial and velocity dimensions. The different features predominantly come from the series of filamentary structures which are attached to the main high density central clump. At $\sim$-50\,pc there is a feature that might even be interpreted as a broad bridge, however this is quite misleading given the scale of the model, as there is actually a spatial offset of  $\sim$10\,pc between the two intensity peaks. Further inspection of the full data cube (and Figure \ref{enzohydro}) reveals this these separated velocity features are due to \textit{multiple} filaments almost coincident along the line of sight. 

This system highlights that p--v diagrams require coordination with other techniques to robustly interpret observations. It also demonstrates that at larger spatial scales interpretation will be more difficult since there will be a greater combination of processes contributing to the emission. Ideally one wants to observe on the scale of the bulk dynamics of one specific process (such as feedback or a collision) with only underlying turbulence as an additional factor, in addition to the larger scale observations.

\section{Comparison with observations of M20}
We find that the broad bridge feature is a signature of cloud--cloud collision that does not arise due to radiative feedback or for clouds merely coincident along the line of sight. Identifying such a feature towards regions of star formation (in particular massive star formation where radiative feedback is prevalent) may therefore provide evidence of star formation triggered by cloud--cloud collision. {In this section we will discuss $^{12}$CO\,J=1-0 observations towards M20 taken with Mopra, which show candidate broad bridge features.}

\subsection{Details of the observations}
{Our $^{12}$CO J=1-0 observations towards M20 were carried out using the 22m ATNF Mopra telescope in Australia in 2011 October. The Mopra backend system ``MOPS'' provided 4096 channels across 137.5 MHz in each of the two orthogonal polarizations, and the corresponding velocity resolution and velocity coverage were 0.088 km/s and 360 km/s, respectively, at the frequency of $^{12}$CO J=1-0 115 GHz. The OTF (on-the-fly) mode was used, and the pointing accuracy was checked every OTF scan to be better than 7\arcsec with SiO maser observations at 86 GHz. The typical system temperature was about 500 K at 115 GHz. The absolute intensity calibration were made with the observations of Orion-KL (R.A., Dec.) =  (5:35:14.5, -5:22:29.6) by comparing the results of \cite{2005PASA...22...62L}. The obtained spectra were gridded to a 15\arcsec spacing and then were spatially smoothed to a beam size 45\arcsec. The achieved rms noise level is $\sim$0.2\,K per channel at velocity resolution 0.9\,km/s.}

\begin{figure*}
	\includegraphics{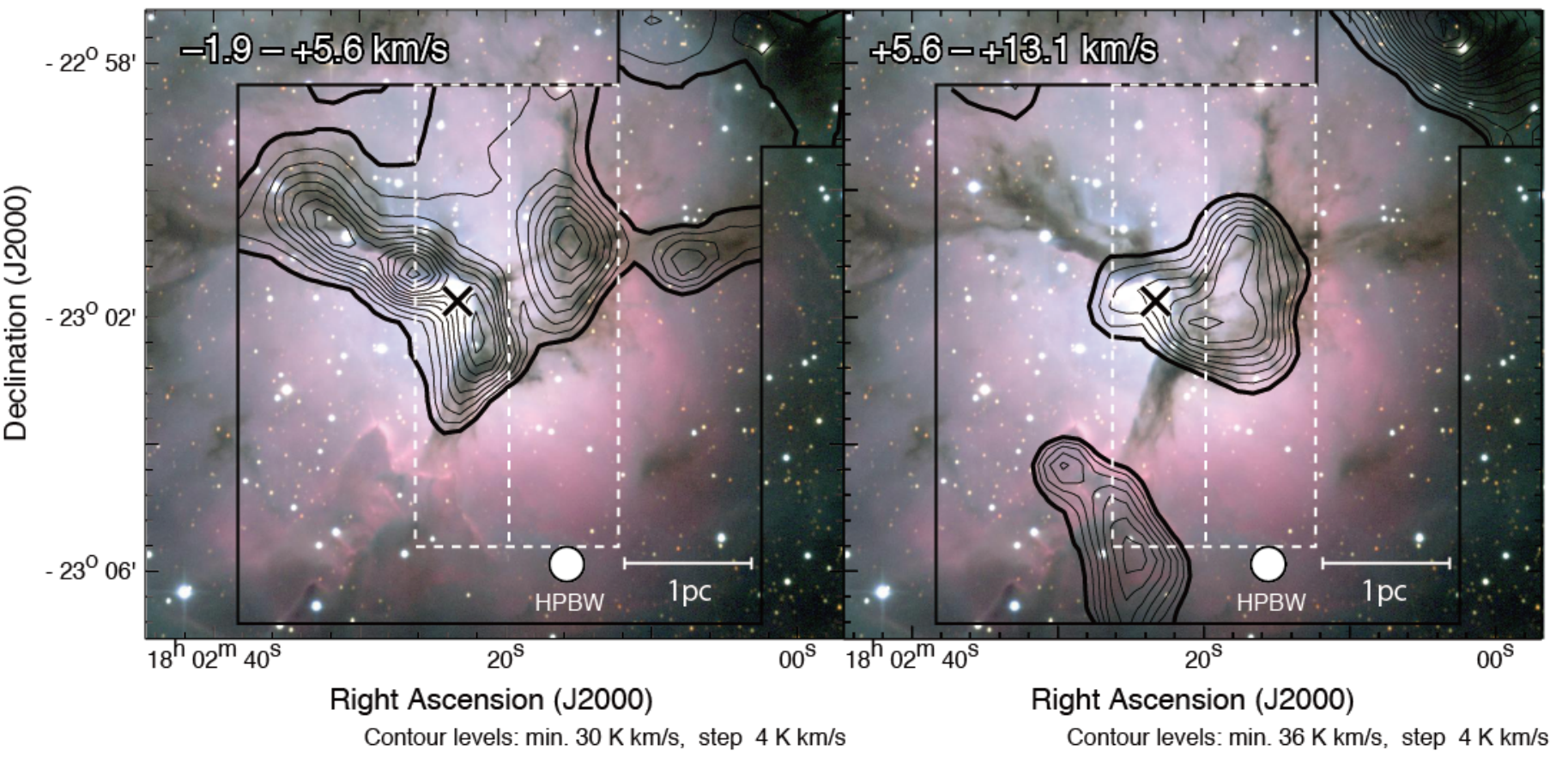}
	\caption{{The two colliding clouds identified by Torii et al. (2011) are presented with the Mopra $^{12}$CO J=1-0 data at 45\arcsec resolution. The background is an optical image of M20 (Credit: Todd Boroson/NOAO/AURA/NSF). The cross indicates the existing O star, responsible for the ionising radiation field. Dashed lines show the regions used for the position-velocity diagrams in Figure \ref{pvmaps_obs}.}}
	\label{M20obs}
\end{figure*}

\subsection{Discussion of M20 observations}

{
Integrated $^{12}$CO J=1-0 emission over two velocity ranges are overlaid upon an optical image in Figure \ref{M20obs}. The black cross represents the exciting O star. In this paper we only discuss p--v diagrams over the white boxed regions, the full presentation of the Mopra dataset will be published elsewhere.
}

\begin{figure}			
	\hspace{20pt}
	\includegraphics[width=7cm, height=7cm]{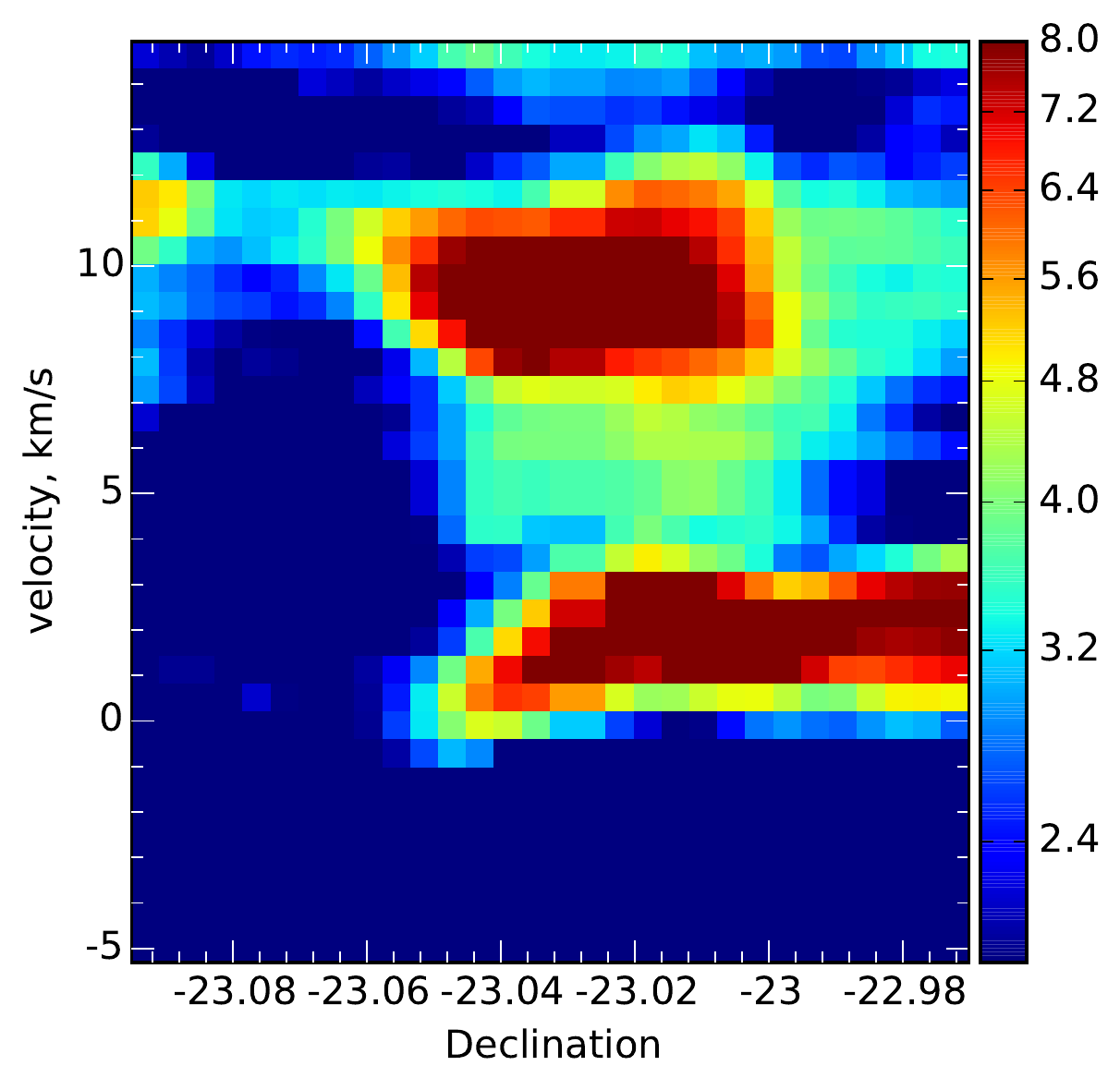}

	\hspace{20pt}	
	\includegraphics[width=7cm, height=7cm]{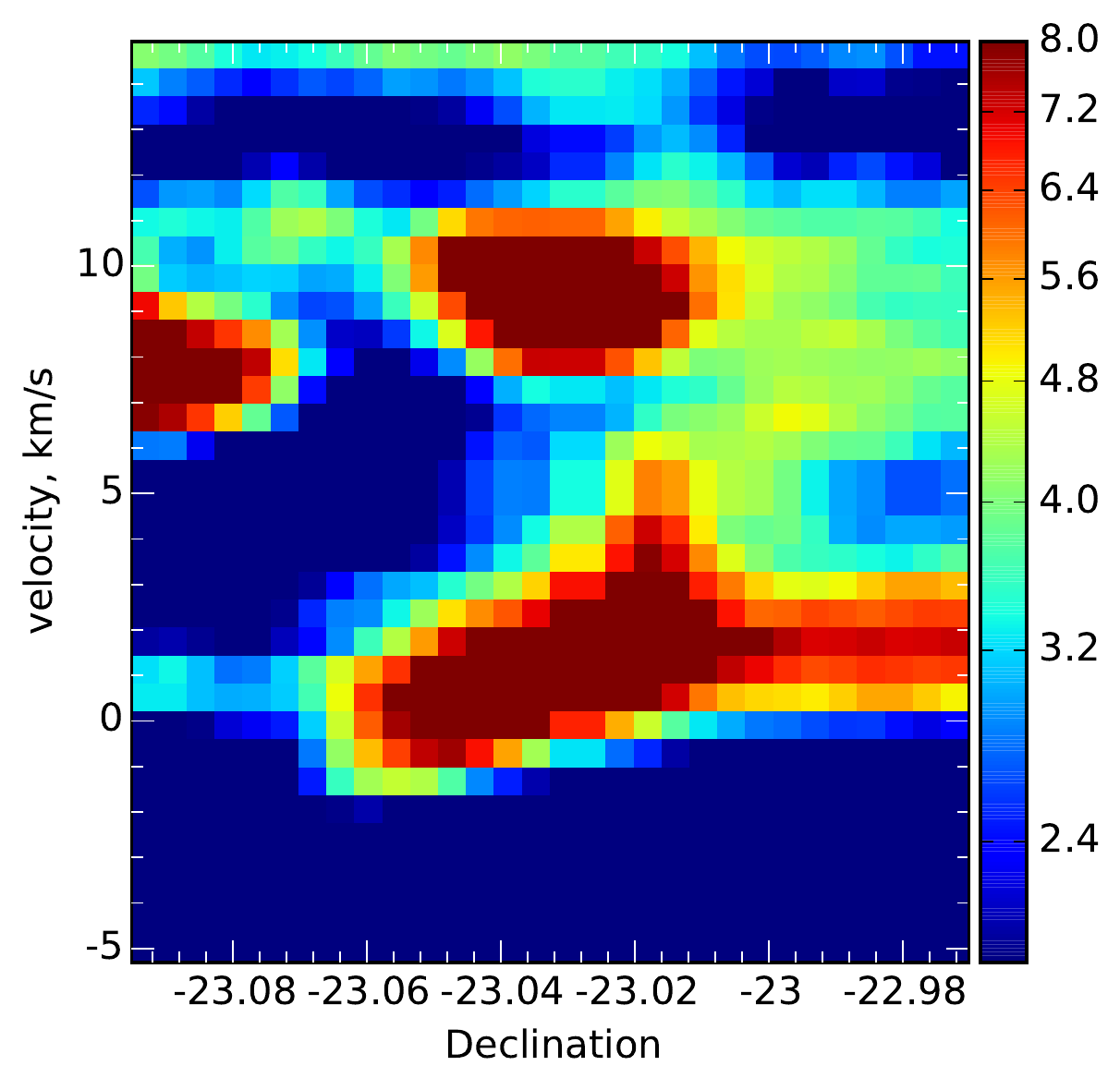}	
	\caption{MOPRA $^{12}$CO J=1-0  position--velocity diagram observations {of candidate broad bridge features} towards M20, a young site of massive star formation. }
	\label{pvmaps_obs}
\end{figure}

Figure \ref{pvmaps_obs}  shows our $^{12}$CO\,J=1-0 position--velocity diagrams towards M20. The upper panel is for gas away from the exciting O7 star  in M20 ({the right hand white boxed region in Figure \ref{M20obs}}) whereas the lower panel is for material close to the star ({the left hand white boxed region in Figure \ref{M20obs}}). The O7 star is at a declination of 23h 031$\arcmin$. Assuming a distance of 1.7\,kpc to M20, both bridge features have a size of roughly a few parsecs (similar to our simulation results). 

There is striking morphological similarity between the observations in Figure \ref{pvmaps_obs} and our simulations in Figure \ref{pvmaps_col}, {though in the lower panel of Figure \ref{pvmaps_obs}  the intermediate velocity gas looks like it is at least partially spatially offset. This could imply chance spatial correlation between three velocity components, or might be due to the action of the ionising radiation field from the O-star disrupting the collision signature.}

M20 is a very young site of massive star formation that was concluded to be the site of a cloud--cloud collision by \cite{2011ApJ...738...46T}. The numerical models in this paper coupled with these MOPRA observations support this hypothesis. Given that this system is very young, we speculate that either the signature of the collision has not yet been disrupted by feedback ({though there could be signs of feedback in the broad bridge close to the O star}), or that the collision is ongoing between different components of the parent molecular clouds. 

A broad bridge feature was also recently identified in p--v diagrams towards star formation in the extreme outer galaxy by \cite{2014ApJ...795...66I}, which they propose was triggered by a high velocity collision.  Our simulations support their conclusion. \cite{2010ApJ...725...17G} also identify  broad bridges in the W33A high mass star forming region, which they interpret in terms of converging flows. 

In general broad bridge features can only really be used in conjunction with other diagnostics. For example, collisions between extended clouds should result in undisturbed subsets of the clouds that will be observable as spatially segregated red and blue shifted gas  \citep[e.g.][]{2011ApJ...738...46T, 2014ApJ...780...36F}

\section{Viewing angle sensitivity}
\label{angles}
So far the discussion of p--v diagrams for collisional models is all regarding a viewing angle along the collision axis ($\theta$=$\pi/2$, $\phi$=0 in Figure \ref{schematic}). 

For the simple geometry of a head on collision we can make a rough estimate of the fraction of viewing angles that a broad bridge is discernible. During the collision the clouds are travelling at velocities $v_1$ and $v_2$ (which may be slower than the pre-collision velocity due to braking) with turbulent velocity dispersions $\Delta v_1$ and $\Delta v_2$. If we require the clouds to be separated by more than their velocity dispersions then simple geometric consideration under the convention given by Figure \ref{schematic} yields
\begin{equation}
	(\Delta  v_1 + \Delta v_2)/2 <  \sin(\theta)\cos(\phi)(v_1 - v_2)
\end{equation}
as the criterion for identifying the broad bridge in a p--v diagram. Note that we have assumed that the velocity dispersion is independent of viewing angle. Integrating over $\theta$ and $\phi$ using $\Delta  v_1=1.71$, $\Delta  v_2=1.25$ (c.f. section \ref{enzo}) and $v_1-v_2\sim3.5$\,km/s (by visual inspection of the middle panel of Figure \ref{pvmaps_col}) gives an estimate of 28 per cent of viewing angles over which the 10\,km/s collision (braked to 3.5\,km/s) will be identified. 

\begin{figure}
\hspace{-25pt}
\includegraphics[width=10cm]{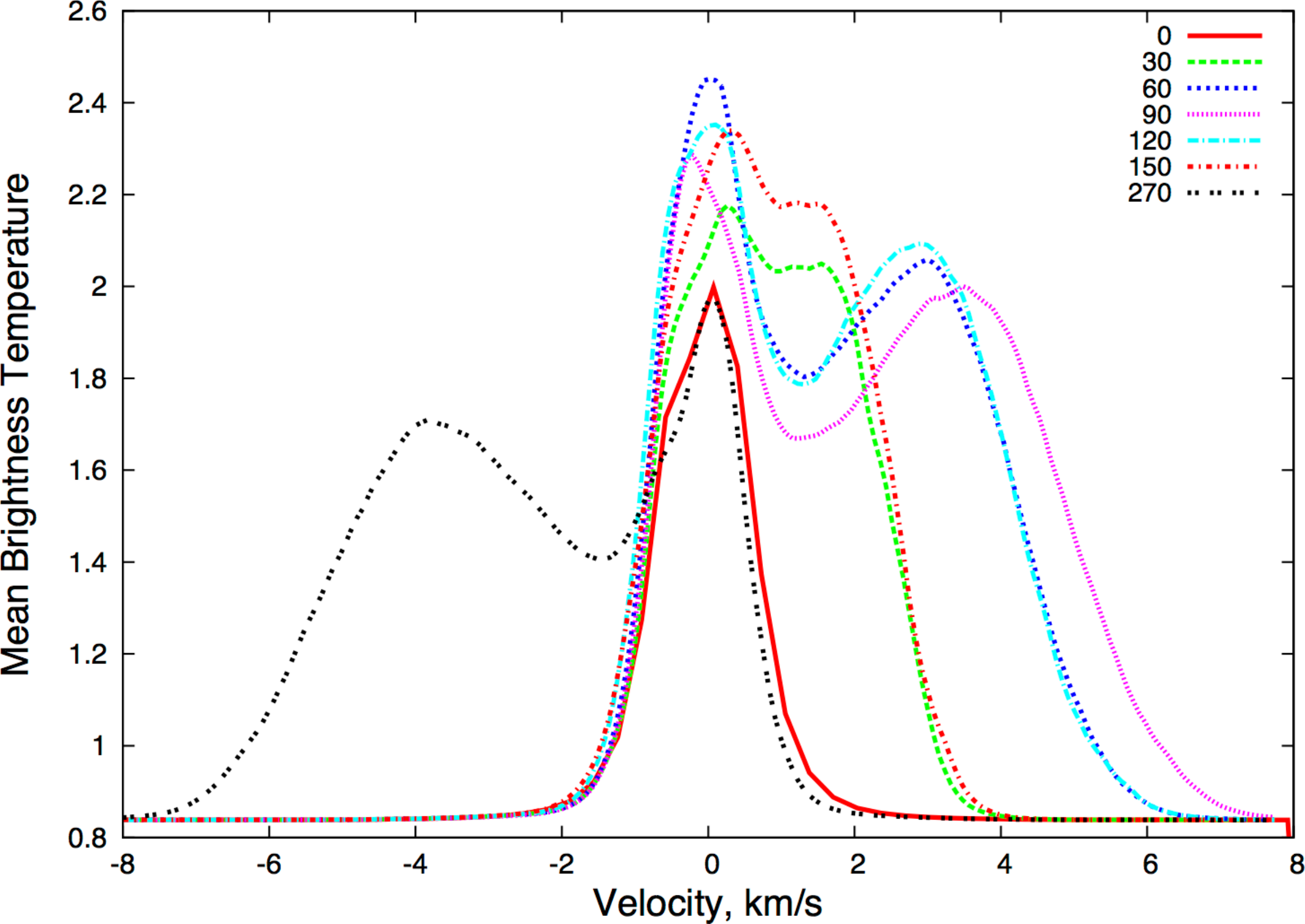}

\hspace{-25pt}
\includegraphics[width=10cm]{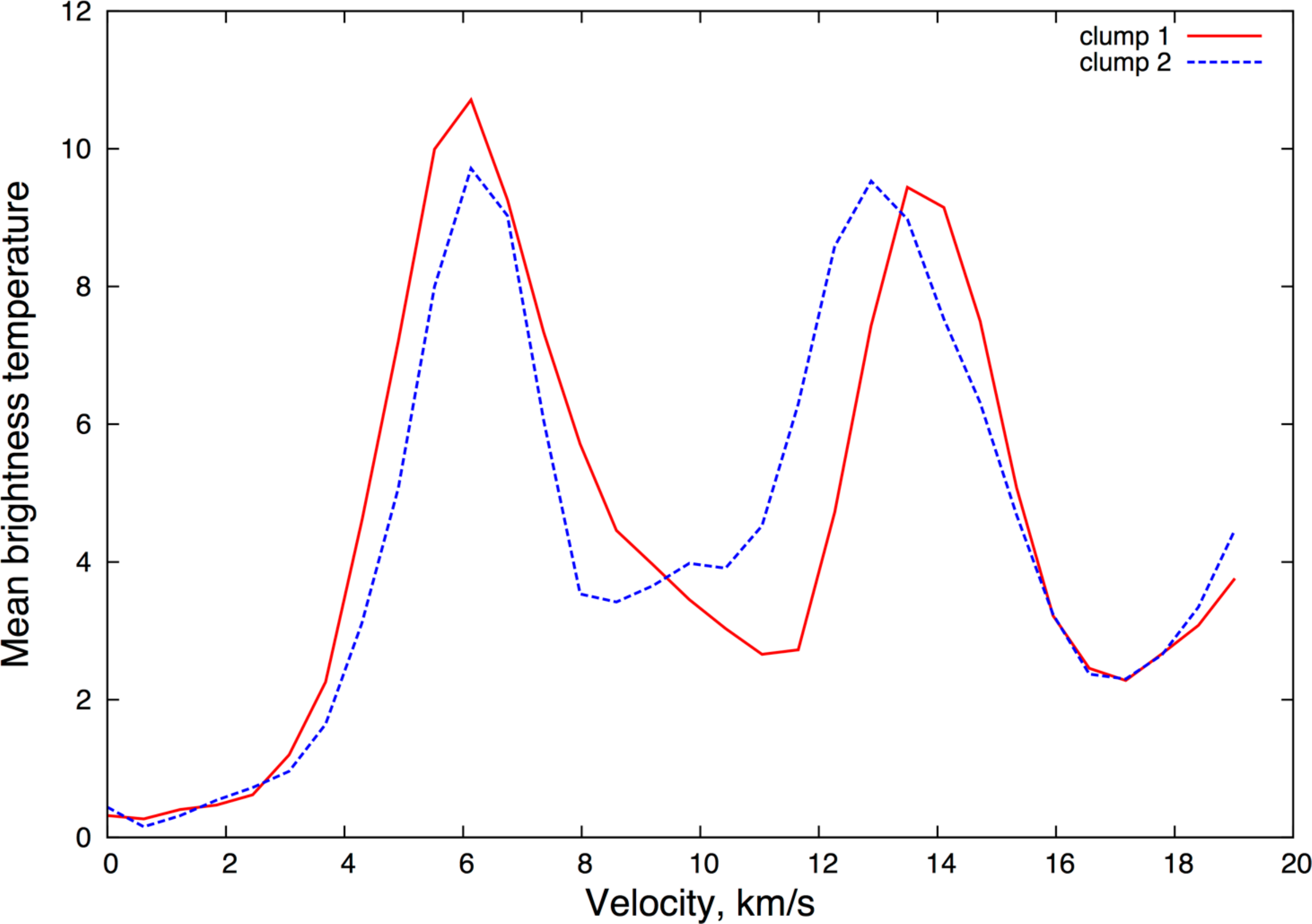}
\caption{The upper panel shows synthetic P-V diagram profiles averaged at each velocity, illustrating the presence of two peaks at some, but not all, viewing angles (where viewing angles are specified in degrees using the convention of Figure \ref{schematic} where $\phi=0$ and $\theta$ varies). Note that this is just an illustrative selection of viewing angles and the profile is reasonable symmetric when the viewing angle is translated by 180  degrees. The lower panel shows the same averaging over the two broad bridges observed towards M20 (clump 1 and 2 are the two regions in the upper and low panel of Figure \ref{pvmaps_obs} respectively).}
\label{meanPV}
\end{figure}

To check our simple estimate we produce p--v diagrams over a range of different viewing angles. To summarise the results of this process we calculate the mean profile over the middle of the p--v diagram where the bridge feature is found. For the 10\,km/s collision model this is shown for a selection of example $\theta$ in the top panel of Figure \ref{meanPV}. At some viewing angles two peaks in the p-v diagram are discernible and we conclude that the broad bridge can be identified, whereas at others there is a single (albeit broad) peak and the broad bridge is not identified. We create a grid of viewing angles spanning $\theta = 0$ to 90 in 10 degree intervals and $\phi = 0$ to 180 in 20 degree intervals and assume that the symmetry of the problem allows us to translate to other viewing angles. We then assume that if there are two peaks in the averaged line profile then the broad bridge (and hence the collision) is identified. Using this definition and spanning our grid we estimate that only 20-30 per cent of viewing angles over $4\pi$ steradians will observe the broad bridge feature in the p--v diagram (at least at this collision velocity and geometry of a head on collision), consistent with our simple estimate. Observing a broad bridge feature might therefore provide evidence for a collision, but because they are only identified from a small range of viewing angles not observing one is not strong evidence for a lack of collision. Conversely, should broad bridge features become commonly identified, this would imply either that collisions take place at velocities much greater than the turbulent velocity (and so are observable at more viewing angles) or there is some further mechanism at work that also gives rise to this signature, making it degenerate.

Note that these conclusions regarding the fraction of viewing angles from which the collision is visible are specific to the geometry of head on colliding clouds with no extended structure that does not undergo collision. It is also specific to the turbulent spectrum of the clouds. Collisions between clouds with lower turbulent velocities will be easier to identify. 

The bottom panel of Figure \ref{meanPV} shows the mean profile over the middle of the p--v diagram for the observations towards M20, clearly showing two discernible peaks similar to those that we see in the models. The M20 peaks are separated by a larger velocity than our simulations, which could mean that the collision was between higher velocity clouds or that it is at an earlier stage than our snapshot times and has not undergone much braking. If the latter, given that stars have already formed, it is possible that the collision is taking place sequentially along different components of the two clouds.

\begin{figure}
	\hspace{-10pt}
	\includegraphics[width=9cm]{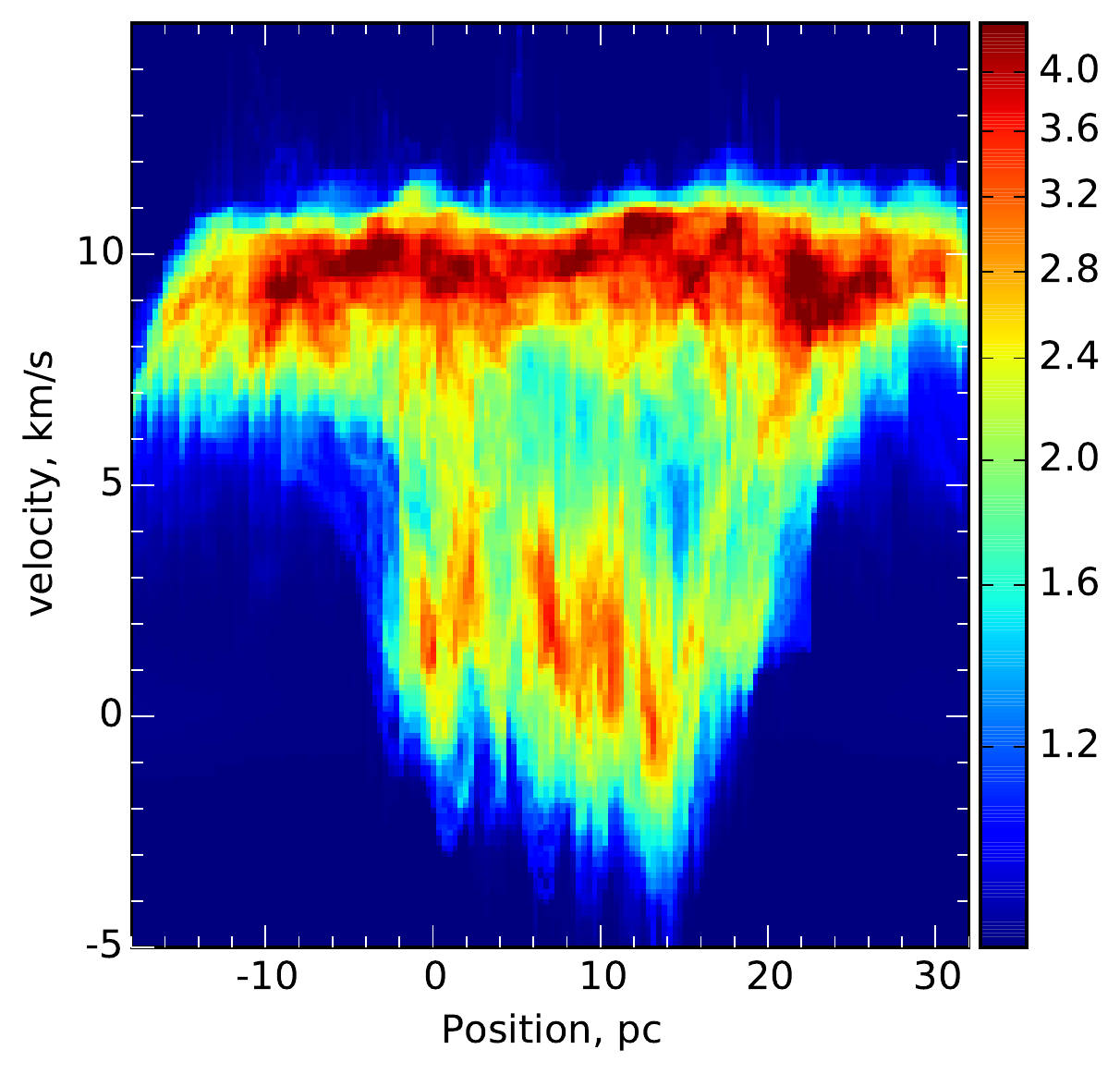}
	\caption{{A $^{12}$CO  J=1-0 position--velocity diagram of a new \textsc{enzo} model of a cloud--cloud collision including star formation and radiative feedback, to be published in Shima et al (in prep). At the time in the simulation that this diagram is produced there are many ionised bubbles due to radiative feedback. This illustrates that the broad bridge is resilient to the effects of radiative feedback. }}
	\label{feedback}
\end{figure}

\section{Model limitations}
\label{limits}
The cloud--cloud collision models considered two isolated, initially spherical, turbulent clouds undergoing a head on collision. In reality these initial conditions are not necessarily representative of those formed in galaxy scale models \citep[e.g.][]{2015MNRAS.446L..46R}. It would be useful to perform a similar analysis using two clouds derived from galaxy scale models, and using different impact parameters. Even better would be to re-run collisions which happen ``naturally'' in galactic scale models at higher resolution.  Ideally future simulations would also model the formation of any star(s) using sink particles. This would also allow us to search for segregated red and blue--shifted cloud components about the star cluster, which are interpreted as evidence of cloud--cloud collision by \citet[e.g.][]{2009ApJ...696L.115F, 2011ApJ...738...46T, 2014ApJ...780...36F}. 

 We have also not determined how long this signature survives after the collision.  {Given that internal radiative feedback seems to be primarily accelerating small clumps (see section \ref{radf}), it is conceivable that the broad bridge might be long lived. To gain an initial insight into the effect of radiative feedback when it comes to disrupting the broad bridge we postprocess a snapshot from the ongoing simulations of Shima et al (in prep). These are similar to the cloud--cloud collision models in this paper, but the clouds are larger, colliding at 20km/s and the simulations also follow the formation of stars and include radiative feedback. For now, we study one snapshot 3\,Myr after the onset of collision and 2.5\,Myr after the formation of the first massive star, at which time there are multiple ionised bubbles. We plot the p--v diagram of this snapshot in Figure \ref{feedback} in which a broad bridge is definitely discernible. At least in this scenario, at this snapshot in time, the broad bridge is still observable despite radiative feedback. We will study the time evolution of the broad bridge in simulations with radiative feedback in subsequent work. }

{ Since we are working on the premise that the cloud collision triggers the formation of massive stars, it should be unlikely that a site of cloud-cloud collision will be subject to an external ionising radiation field unless the collision is sequential along the cloud length (i.e. the collision is ongoing after the formation of massive stars) which is not the case in the collision models studied here. }

Finally, the astrophysical scenarios that we explore may not be exhaustive. For example, we do not investigate p--v diagrams for clouds disrupted by supernovae, large scale flows due to gravity, {radiative feedback in less leaky} H\,\textsc{ii} {regions (which should give the elliptical signature discussed in section \ref{radf})} or those traversing spiral arm shocks. In particular outflows can produce high velocity features, which we do not study here. If any of these processes can also give rise to a broad bridge then the signature becomes degenerate (though potentially still useful in conjunction with other diagnostics).

\section{Summary and conclusions}
We have calculated synthetic p--v diagrams for a number of different astrophysical systems: cloud--cloud collisions, non--interacting turbulent clouds coincident along the line of sight, turbulent clouds with internal radiative feedback and a GMC  evolving in a galactic disc. We compare these to try and identify characteristic signatures of cloud--cloud collisions.  We draw the following conclusions from this work:\\

\noindent 1) Cloud--cloud collision models give rise to a broad bridge structure in p--v diagrams. This broad bridge is two intensity spikes separated by lower intensity emission across the velocity axis. This feature is not reproduced by p--v diagrams from any of the other scenarios, potentially making it a useful signature for identifying cloud--cloud collisions.  We also find instances of this broad bridge feature towards M20, a very young site of massive star formation which was concluded to be a site of cloud--cloud collision by \cite{2011ApJ...738...46T}. Our models therefore support the conclusion that M20 is a site of cloud--cloud collision. \\


\noindent 2) The broad bridge feature is observable in our 10\,km/s collision model, but not in our 3\,km/s model.  We conclude that in order to observe the broad bridge, the difference in line of sight velocity between the two colliding clouds at the time of observation has to be greater than half the sum of the turbulent velocity dispersions of the clouds.  \\

\noindent 3) Using the criterion from conclusion 2, we estimate that for the specific case of our 10\,km/s head on collision model, the broad bridge is only observable over 20-30 per cent of $4\pi$ steradians. Given that it is also not clear how long the broad bridge survives before being disrupted by star formation and/or feedback, not observing a broad bridge can therefore not be used to rule out cloud-cloud collisions. Conversely widespread identification of this feature might suggest low turbulence, fast collisions, the importance of impact parameter, or alternatively that some other mechanism also gives rise to a broad bridge.\\

\noindent 4) {Using ongoing models from Shima et al (in prep), we preliminarily show that the broad bridge feature is resilient to the effects of radiative feedback at least up to 2.5\,Myr after the formation of the first massive star.} \\

\noindent 5) In general, a broad bridge feature seems to provide a useful signature of cloud--cloud collision, but should be used in conjunction with other diagnostics such as channel maps and checking for high velocity segregated blue and red shifted clouds near the proposed collision site that may be relics of the pre--collision clouds. \\

\section*{Acknowledgements}
TJH is funded by the STFC consolidated grant ST/K000985/1. EJT is funded by the MEXT grant for the Tenure Track System. {Enzo simulations were carried out on the Cray XC30 at the Center for Computational Astrophysics (CfCA) of the National Astronomical Observatory of Japan. Visualisations were done using \textsc{yt} \citep{2011ApJS..192....9T}}. {We thank the anonymous referee for their thorough and insightful review, which improved the manuscript}. We also thank Clare Dobbs and Ana Duarte-Cabral for their comments on an earlier draft of the manuscript. This project was initiated following an informal meeting on triggered star formation hosted at University College London in 2014 and organised by Thomas Bisbas.
This work used the DiRAC Data Analytics system at the University of Cambridge, operated by the University of Cambridge High Performance Computing Serve on behalf of the STFC DiRAC HPC Facility (www.dirac.ac.uk). This equipment was funded by BIS National E-infrastructure capital grant (ST/K001590/1), STFC capital grants ST/H008861/1 and ST/H00887X/1, and STFC DiRAC Operations grant ST/K00333X/1. DiRAC is part of the National E-Infrastructure. This research was supported by the DFG cluster of excellence `Origin and Structure of the Universe (JED)'.

\bibliographystyle{mn2e}
\bibliography{molecular}

\bsp

\label{lastpage}

\end{document}